\documentclass[reprint,prx,footinbib,floatfix,twocolumn,longbibliography,superscriptaddress]{revtex4-2}
\usepackage{times}
\usepackage{amssymb}
\usepackage{color}
\usepackage{amsmath}
\usepackage{amsbsy}
\usepackage{amsthm}
\usepackage{graphicx}
\usepackage{bbm}
\usepackage{bm}
\usepackage{epsfig}
\usepackage{xfrac}
\usepackage{xcolor}
\usepackage{enumerate}
\usepackage{multirow}
\usepackage{physics}
\usepackage[T2A,T1]{fontenc}
\usepackage{ tipa }
\usepackage[english]{babel}
\usepackage{symbols}
\usepackage{makecell}
\usepackage{appendix}
\usepackage{dsfont}
\usepackage{float}
\usepackage{pdfpages}
\usepackage{bbold}
\usepackage{relsize}
\usepackage{xcolor}
\usepackage{scalerel}
\usepackage{tikz}
\usetikzlibrary{svg.path}
\usepackage{stackengine}
\usepackage{multirow}
\usepackage{enumitem}

\definecolor{orcidlogocol}{HTML}{A6CE39}
\tikzset{
  orcidlogo/.pic={
    \fill[orcidlogocol] svg{M256,128c0,70.7-57.3,128-128,128C57.3,256,0,198.7,0,128C0,57.3,57.3,0,128,0C198.7,0,256,57.3,256,128z};
    \fill[white] svg{M86.3,186.2H70.9V79.1h15.4v48.4V186.2z}
                 svg{M108.9,79.1h41.6c39.6,0,57,28.3,57,53.6c0,27.5-21.5,53.6-56.8,53.6h-41.8V79.1z M124.3,172.4h24.5c34.9,0,42.9-26.5,42.9-39.7c0-21.5-13.7-39.7-43.7-39.7h-23.7V172.4z}
                 svg{M88.7,56.8c0,5.5-4.5,10.1-10.1,10.1c-5.6,0-10.1-4.6-10.1-10.1c0-5.6,4.5-10.1,10.1-10.1C84.2,46.7,88.7,51.3,88.7,56.8z};
  }
}

\newcommand\orcid[1]{\href{https://orcid.org/#1}{$\,$\mbox{\scalerel*{
\begin{tikzpicture}[yscale=-1,transform shape]
\pic{orcidlogo};
\end{tikzpicture}
}{|}}}}

\definecolor{myurlcolor}{rgb}{0.0,0.39,0.0}
\definecolor{myrefcolor}{rgb}{0.0,0.39,0.0}

\usepackage[colorlinks]{hyperref}
\hypersetup{unicode=true,
    bookmarksopen=false,
    breaklinks=false,
    pdfborder={0 0 0},
	bookmarksnumbered=false,
	pdfstartview={FitH},
	citecolor={myurlcolor},
	linkcolor={myrefcolor},
	urlcolor={myurlcolor}}

\definecolor{cyan(process)}{rgb}{0.0, 0.72, 0.92}

\makeatletter
\AtBeginDocument{\let\LS@rot\@undefined}
\makeatother

\begin{document}

\title{Quantum Non-Gaussian State Preparation of Levitated Particles via Time-Dependent Control of Weakly Nonharmonic Hybrid Potentials}

\author{Piotr~T.~Grochowski\orcid{0000-0002-9654-4824}}
\email{piotr.grochowski@upol.cz}
\affiliation{Department of Optics, \href{https://ror.org/04qxnmv42}{Palacký University}, 17. listopadu 1192/12, 771 46 Olomouc, Czech Republic}

\author{Oriol~Romero-Isart\orcid{0000-0003-4006-3391}}
\email{oriol.romero-isart@icfo.eu}
 \affiliation{\href{https://ror.org/03g5ew477}{ICFO - Institut de Ciencies Fotoniques, The Barcelona Institute of Science and Technology}, 08860 Castelldefels (Barcelona), Spain}
 \affiliation{\href{https://ror.org/0371hy230}{ICREA - Institucio Catalana de Recerca i Estudis Avançats}, 08010 Barcelona, Spain}

\begin{abstract}
Levitated high-mass quantum systems provide access to unprecedented regimes in both fundamental science and technological applications.
However, deterministic generation and manipulation of quantum non-Gaussian states, which are central to many continuous-variable quantum advantages, remain elusive in such platforms.
In this work, we propose a theoretical protocol for preparing a continuous-variable degree of freedom of a levitated massive object in a variety of quantum states, including Fock and Schr\"odinger cat states, without coupling to auxiliary two-level systems.
Our approach enhances otherwise weak nonharmonic effects by transient wave-function delocalization and combines this with optimal control of the potential.
Specifically, time-dependent modulation of the linear component of the potential, in the presence of a static cubic nonharmonicity, provides a route to universal control of the mode.
We analyze quantum state preparation under such control and estimate the required nonharmonicity, motional delocalization, and maximum tolerable decoherence for generating target non-Gaussian states.
The proposed optimal-control scheme can also be readily extended beyond single-particle state preparation, for example, to unitary transformations and nonlinear measurements.
As a concrete example, we demonstrate mechanical Bell-state preparation for two interacting particles using only local modulation of weakly nonharmonic potentials, while the interparticle interaction remains effectively linear.
We emphasize that the protocols presented here apply to different mechanical degrees of freedom, such as center-of-mass motion and libration, and can also be implemented in other weakly nonharmonic systems with a leading cubic nonharmonicity.
\end{abstract}
\maketitle

\section{Introduction}
In recent years, significant effort has focused on controlling quantum systems in extreme parameter regimes, including large spatial scales~\cite{Manetsch2025}, highly energetic motion~\cite{TheATLASCollaboration2024}, and very massive objects~\cite{Whittle2021}, to probe new physics and enable technological applications.
In particular, the motion of massive systems enables investigations at the quantum–gravity interface~\cite{Bose2025}, tests of modifications of the Schr\"odinger equation~\cite{Romero-Isart2011a}, and searches for dark matter~\cite{Higgins2024}.
From a technological perspective, these systems offer potential applications as ultraprecise inertial sensors~\cite{Geraci2015,Hebestreit2018,Weiss2021} and frequency transducers~\cite{Bagci2014,Andrews2014,Forsch2020}.
Many routes to quantum advantage in these regimes, however, require the preparation and manipulation of large-scale, high-quality non-Gaussian states~\cite{Walschaers2021,Rakhubovsky2024,Fadel2025,Grochowski2025}.

\begin{figure}[ht!]
    \includegraphics[width=\linewidth]{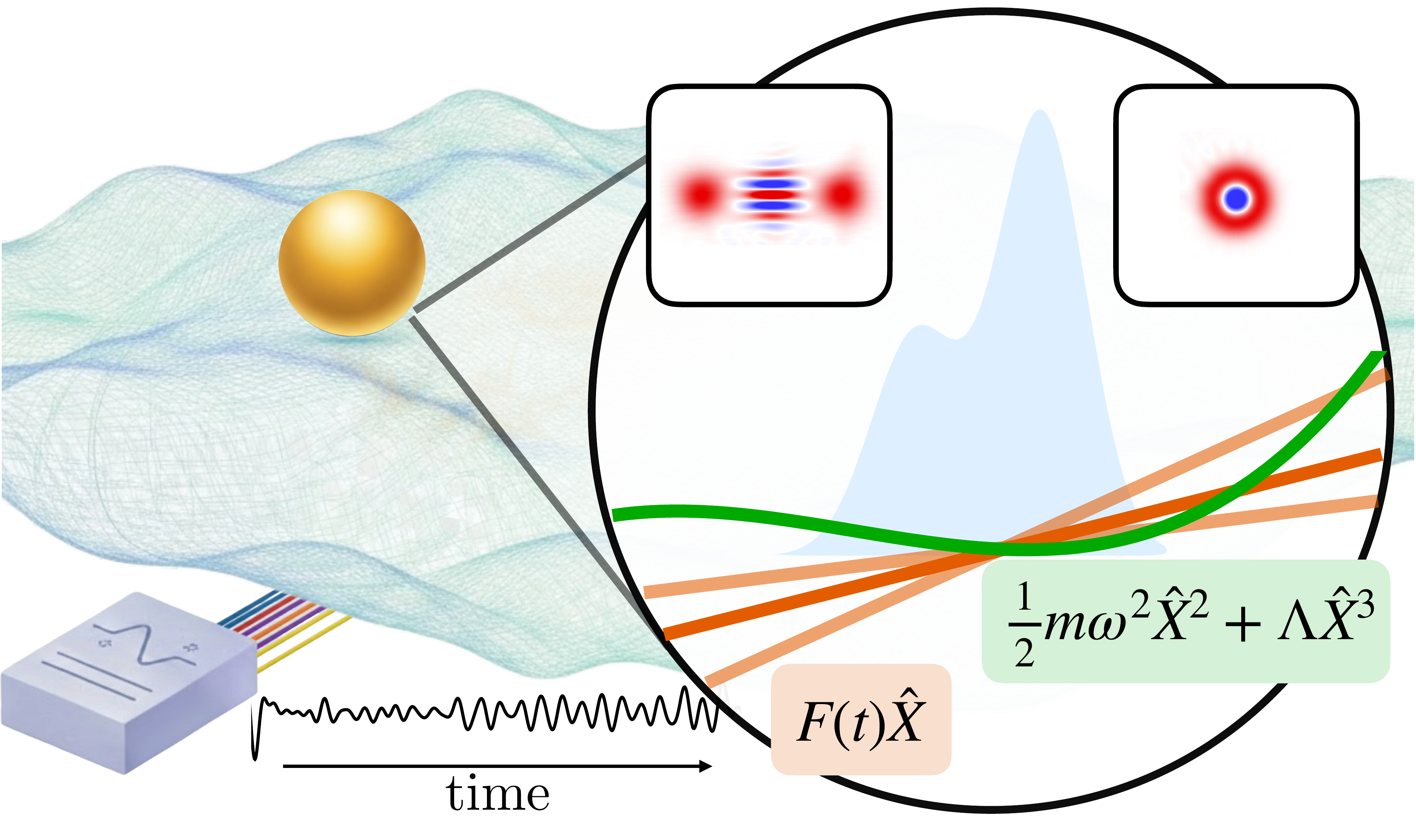}
    \caption{ We consider a levitated mechanical oscillator whose mechanical state---for example, its center-of-mass motion or libration---evolves in a weakly nonharmonic potential landscape, $V(\PositionOperator) = \Mass \LowFrequency^2 \PositionOperator^2/2 + \CubNonlinDim \PositionOperator^3$. We demonstrate that temporal modulation of such a potential, where the leading nonquadratic contribution is cubic, $\CubNonlinDim \PositionOperator^3$, enables deterministic preparation of quantum non-Gaussian states such as Fock and Schr\"odinger cat states. This scheme exploits the enhancement of nonlinear dynamics arising from wave-function delocalization, combined with optimal control of the modulation.
    Specifically, we control the local linear term, $F(\DimTime) \PositionOperator$, while keeping the nonharmonicity static.
    The proposed state-preparation protocol can be extended to scenarios involving additional degrees of freedom or multiple oscillators, enabling, for example, Bell-state generation.
    Our approach can also be used in other quantum systems with weak cubic nonlinearities, avoiding the need for auxiliary nonlinear elements.\label{Fig1}}
\end{figure}

While non-Gaussian control in the quantum regime has been achieved in some clamped mechanical systems~\cite{Bild2023,Marti2024,Galinskiy2024,Yang2024}, it remains unrealized for levitated mechanical oscillators~\cite{Gonzalez-Ballestero2021,Millen2020}, which, in contrast, offer excellent environmental isolation and flexible potential shaping~\cite{Rashid2016,Hebestreit2018,Frimmer2019,Ciampini2025,Bonvin2024}.
Ground-state cooling of a high-mass levitated system has been demonstrated with silica nanoobjects trapped in optical potentials, for both translational~\cite{Delic2020,Magrini2021,Tebbenjohanns2021,Kamba2022,Ranfagni2022,Piotrowski2023,Kamba2023a,Deplano2025} and librational~\cite{Dania2025,Troyer2026} degrees of freedom.
Beyond cooling, proposals and experimental efforts have aimed to create non-Gaussian states through coherent coupling to two-level systems~\cite{Romero-Isart2012,Hoang2016,Martinetz2020,Perdriat2021,Marshman2022,Bykov2025,Gupta2025}, measurement-induced heralding~\cite{Bemani2025}, or potential engineering alone.
The last strategy uses wave-function delocalization~\cite{Romero-Isart2017,Weiss2021,Braccini2024,Zhou2025,Ji2026} to enhance the effect of intrinsically weak trapping-potential nonharmonicities~\cite{Riera-Campeny2024,Roda-Llordes2024b,Rosiek2024}, whose influence is otherwise suppressed at the zero-point length scale.
Recent experiments have aimed to achieve such state expansion for initially thermal~\cite{Rashid2016,Hebestreit2018,Kamba2023,Bonvin2024,Muffato2025,Duchan2025,Steiner2025,Tomassi2026,Seta2026} and near-ground-state~\cite{Rossi2025,Kamba2025,Otabe2026,Skrabulis2026} motion; only in the latter four studies was the motion squeezed below the zero-point fluctuations.
More specifically, potential-engineering proposals aim to generate cubic-phase~\cite{Gottesman2001,Rakhubovsky2021,Neumeier2024,Roda-Llordes2024,Casulleras2024} or quartic-phase~\cite{Rosiek2024} states by exploiting nonlinear dynamics in nonoptical potentials, thereby avoiding photon-scattering and optical-noise decoherence associated with optical trapping~\cite{Jain2016,Pino2018,Maurer2023}.
Related deterministic strategies based on nonlinear blockade dynamics in engineered oscillator potentials have also been proposed~\cite{Huang2021}.

While existing proposals focus on generating and maximizing verifiable quantum non-Gaussian features~\cite{Neumeier2024,Roda-Llordes2024b,Riera-Campeny2026}, a framework for universal control of a levitated particle’s mechanical degree of freedom remains lacking.
Such control would, among other applications, enable tests of gravity with massive oscillators prepared in non-Gaussian motional states, enhance studies of quantum gravitational coupling in the low-energy limit, and help constrain semiclassical and collapse-type models~\cite{Romero-Isart2011,Bose2025}.
At the same time, deterministic access to non-Gaussian mechanical resources would enable the use of fine phase-space structure beyond Gaussian limits~\cite{Fadel2025,Grochowski2025,Park2025}, supporting quantum-enhanced force and acceleration sensing, including in distributed schemes~\cite{Grochowski2026}.

Here, we address this gap by extending potential-engineering approaches to include time-dependent potential modulation, providing a route to universal control (cf. Fig.~\ref{Fig1}).
We harness well-controlled nonlinear dynamics induced by the leading nonquadratic term of the confining potential, which can be cubic when the expansion point lacks inversion symmetry~\cite{Rakhubovsky2021,Neumeier2024}.
Because this nonharmonicity is extremely weak for typical levitated massive objects, we enhance its effect through wave-function delocalization.
As the central control knob, we use an optimally modulated~\cite{Grochowski2025a,Kendell2024} linear component of the potential, analogous to classical linear drives in platforms such as trapped ions~\cite{Leibfried2003a} and cavity or circuit quantum electrodynamics~\cite{Schuster2005,Devoret2013}.
Crucially, we utilize strong wave-function delocalization only as an intermediate resource that transiently enhances the otherwise weakly nonharmonic dynamics~\cite{Rosiek2024}.
The transiently delocalized state is subsequently coherently compressed back to the tightly confined trapping potential, where it is more robust against decoherence and directly accessible to optical manipulation and measurement.

Such a controlled potential landscape allows, in principle, the generation and manipulation of a broad class of quantum non-Gaussian states.
We focus on high-occupation Fock states and large-amplitude Schrödinger cat states as representative examples.
We identify relevant regimes for such state-preparation protocols, including the minimum required wave-function delocalization for a given available nonharmonicity and the constraints imposed by noise.
Moreover, we extend the proposed protocol to two-mode operations, enabling Bell-state generation either between separated interacting particles or between two mechanical modes of a single levitated object.

The proposed approach can be implemented in hybrid optical–nonoptical setups, where the trapping potential is switched between optical fields used for measurement and electric or magnetic fields used for lower-noise dynamical evolution.
Beyond center-of-mass motion, other continuous degrees of freedom can be described within the same framework, most prominently libration.
Finally, we stress that our approach is not specific to massive levitated oscillators and can be applied to other quantum systems with weak potential nonharmonicities.

The manuscript is structured as follows.
In Section~\ref{sec_mass}, we introduce the model of a massive particle evolving in a nonharmonic potential landscape and discuss wave-function delocalization as a means of enhancing nonlinear dynamics and its tradeoff with decoherence. 
Section~\ref{sec_state-prep} describes the state-preparation protocol in detail, while Section~\ref{sec_nongauss} presents representative realizations for high Fock and cat states, together with an analysis of the relevant scalings.
Section~\ref{sec_exp} identifies and models relevant sources of noise for the state-generation protocol.
Section~\ref{sec_beyond} goes beyond single-particle state preparation by analyzing mechanical Bell-state generation in two-mode systems, namely two Coulomb-interacting particles or two motional modes of a single particle.
Finally, Section~\ref{sec_conc} provides conclusions and outlook.

\section{Massive particle in a nonharmonic potential landscape}
\label{sec_mass}
We consider the motion of a particle of effective mass $\Mass$ with canonical position and momentum operators $\PositionOperator$ and $\MomentumOperator$, such that $[\PositionOperator, \MomentumOperator ] = \ImagUnit \hbar$.
The dynamics of this mode takes place in a nonharmonic external potential landscape $\Pot (\PositionOperator)$, which we expand in a Taylor series around a point $\PosDim_0$.
After shifting the coordinate origin to $\PosDim_0$,
\begin{align}
    \Pot (\PositionOperator) & \approx \DimForce \PositionOperator + \frac{1}{2}\Mass \UnFreq^2 \PositionOperator^2 + \CubNonlinDim \PositionOperator^3 \nonumber \\
    & = \hbar \UnFreq \rounds{\force \PosOp_{\UnFreq} + \frac{1}{2} \PosOp_{\UnFreq}^2 +  \NonLin_{\UnFreq} \PosOp_{\UnFreq}^3 },
    \label{potlandscape}
\end{align}
where $\DimForce = \Pot^{(1)} \rounds{\PosDim_0}$, $\UnFreq = \sqrt{ \Pot^{(2)} \rounds{\PosDim_0} / \Mass}$, $\CubNonlinDim  = \Pot^{(3)} \rounds{\PosDim_0}/6$; we introduce local dimensionless operators, 
\begin{align}
    \PosOp_{\UnFreq} &= \frac{\PositionOperator}{  \PosDim_{\UnFreq}} = \frac{\Creation_{\UnFreq}+\Annihilation_{\UnFreq}}{\sqrt{2}}\nonumber, \ \ \ \MomOp_{\UnFreq} =  \frac{\MomentumOperator   \PosDim_{\UnFreq}} { \hbar }= \ImagUnit  \frac{\Creation_{\UnFreq}-\Annihilation_{\UnFreq}}{\sqrt{2}},
\end{align}
that are defined via oscillator length $\PosDim_{\UnFreq} = (\hbar / \Mass \UnFreq)^{1/2}$ associated with local frequency $\UnFreq$; and we use a shorthand notation $\Pot^{(n)} \rounds{\PosDim_0} \equiv \partial_{\PosDim}^n\Pot(\PosDim)|_{\PosDim = \PosDim_0 }$.
Note that throughout the manuscript we consider three specific trapping frequencies, $\UnFreq \in \{\LowFrequency,\sqrt{\LowFrequency\Frequency},\Frequency\}$, while the symbol $\UnFreq$ is otherwise used as a generic frequency variable.
The second line in Equation~\eqref{potlandscape} gives the potential in harmonic oscillator units associated with the frequency $\UnFreq$.
Moreover, we have introduced a cubic nonharmonicity parameter,
\begin{align}
    \NonLin_{\UnFreq} = \frac{\CubNonlinDim \PosDim_{\UnFreq}^3  }{ \hbar \UnFreq} = \sqrt{\frac{\hbar}{m^3}} \frac{\CubNonlinDim}{\UnFreq^{5/2}}.
    \label{nonlinpar}
\end{align}
Note that $\NonLin_{\UnFreq}$ does not depend on the state, but rather characterizes the potential only.
It locally compares the strengths of the quadratic and cubic parts of the Hamiltonian, and quantifies how much nonharmonicity the ground state of a harmonic oscillator with frequency $\UnFreq$ would experience.
We have truncated the expansion~\eqref{potlandscape} at the leading nonquadratic term, assuming that higher-order terms are negligible over the phase-space region explored by the protocol.
We assume that the couplings to other degrees of freedom are either negligible, act as a decoherence channel, or can be accounted for with a full numerical approach~\cite{Roda-Llordes2024b}.
The degree of freedom given by the operator $\PositionOperator$ can naturally be interpreted as the center-of-mass motion of a particle.
However, the analysis of the dynamics in the nonharmonic potential presented here can be equivalently realized in other degrees of freedom, such as libration.
In such a case, the parameter $\Mass$ can be identified with the moment of inertia of the particle, $\PositionOperator$ with its orientation angle, and $\MomentumOperator$ with its angular momentum.

\subsection{Nonharmonicity enhancement and decoherence evasion via motional delocalization}
Each term in~\eqref{potlandscape} defines a characteristic time scale for the corresponding change in the quantum state~\cite{Riera-Campeny2024}.
Typically, these unitary changes compete with dissipative processes, leading to the noisy evolution,
\begin{align}
    \partial_{\DimTime} \DensityMatrix= \frac{1}{i \hbar} [\Hamiltonian,\DensityMatrix] + \CollapseFockCoeff\Dissipator[\CollapseFock]\DensityMatrix,
    \label{mastereq0}
\end{align}
where the density operator $\DensityMatrix$ describes the state of a particle, the Hamiltonian $\Hamiltonian$ consists of both the kinetic and potential parts, $\Hamiltonian = \MomentumOperator^2 / 2 \Mass +  \Pot (\PositionOperator) $, and the Lindblad dissipator is given by $\Dissipator[\CollapseFock]\DensityMatrix = \CollapseFock \DensityMatrix \CollapseFock^\dagger - (\CollapseFock^\dagger \CollapseFock  \DensityMatrix + \DensityMatrix  \CollapseFock^\dagger \CollapseFock )/2$ with the collapse operator $ \CollapseFock$.
The instantaneous state-change rates can be defined from the short-time fidelity decay~\cite{Bures1969,Jozsa1994}, 
\begin{align}
    \Fidelity[\DensityMatrix(\DimTime),\DensityMatrix(\DimTime + \dd \DimTime) ] &= \bra{\WaveFunction (\DimTime)}\DensityMatrix(\DimTime + \dd \DimTime)  \ket{\WaveFunction (\DimTime)}  \nonumber \\ 
    & \approx 1 - (\Rate_\text{unit} \dd \DimTime )^2 - \Rate_\text{diss} \dd \DimTime
\end{align}
with 
\begin{align}
   \Rate_\text{unit} =  \sqrt{\Var_{\WaveFunction} \Hamiltonian}/\hbar, \ \ 
   \Rate_\text{diss} = \CollapseFockCoeff \Var_{\WaveFunction} \CollapseFock.
\end{align}
Here, we assume that the state $\DensityMatrix(\DimTime)$ is pure~\cite{Jozsa1994},  $\DensityMatrix(\DimTime) = \ket{\WaveFunction (\DimTime)} \bra{\WaveFunction (\DimTime)}$, the collapse operator is Hermitian, $\CollapseFock = \CollapseFock^\dagger$, and $\Var_{\WaveFunction} \hat{A} = \bra{\WaveFunction} (\hat{A}  - \langle \hat{A}  \rangle)^2 \ket{ \WaveFunction}$.
These rates are closely associated with time scales known as quantum speed limits, which have been introduced in both isolated~\cite{Mandelstam1945,Uffink1993,Margolus1998,Levitin2009} and open~\cite{Taddei2013,delCampo2013,Deffner2013a} quantum systems.
The unitary rate is given by the quantum Fisher information~\cite{Braunstein1994} with respect to the time $\DimTime$, $\Rate_\text{unit} = \sqrt{F_\text{Q}(\DensityMatrix)}/2$, and yields the Mandelstam-Tamm bound~\cite{Mandelstam1945}.
The dissipative rate analogously quantifies irreversible distinguishability generation.
In typical regimes, the evolution of levitated nanoparticles is dominated by Gaussian dynamics, generated by the quadratic terms of the Hamiltonian.
For Gaussian states with zero means, $\langle \PositionOperator \rangle = \langle \MomentumOperator \rangle = 0$, we have $\Var_{\WaveFunction} \PositionOperator^k = (2k-1)!! (\Delta \PosDim)^{2k}$ for odd $k$ and $\Var_{\WaveFunction} \PositionOperator^k = [(2k-1)!! - (k-1)!!^2](\Delta \PosDim)^{2k}$ for even $k$ with $\Delta \PosDim = \sqrt{\Var \PositionOperator}$.
Considering only the contribution from the harmonic potential term, $\Hamiltonian_\text{G} = \Mass \UnFreq^2 \PositionOperator^2/ 2$, we obtain the Gaussian evolution rate,
\begin{align}
    \GaussianRate = \frac{ \Mass \UnFreq^2 }{\sqrt{2}\hbar }(\Delta \PosDim)^2  = \frac{\UnFreq}{\sqrt{2}}  \frac{(\Delta \PosDim)^2}{ \PosDim_{\UnFreq}^2},
\end{align}
while the leading nonharmonic term $\Hamiltonian_\text{nG} = \CubNonlinDim \PositionOperator^3$ generates non-Gaussian evolution at the rate
\begin{align}
    \NonGaussianRate = \frac{\sqrt{15}|\CubNonlinDim |}{\hbar} (\Delta \PosDim)^3  = \sqrt{15} \UnFreq  |\NonLin_{\UnFreq}|  \frac{(\Delta \PosDim)^3}{   \PosDim_{\UnFreq}^3 }.
\end{align}
This contribution becomes increasingly important as the wave-function spread grows, $\NonGaussianRate / \GaussianRate \sim |\NonLin_{\UnFreq}| \Delta \PosDim / \PosDim_{\UnFreq}$, highlighting the role of wave-function delocalization in enhancing nonlinear dynamics.
The dominant decoherence mechanism in many levitated systems can be modeled as mechanical diffusion (see Section~\ref{sec_exp}).
We write it as $\DimNoise_{\UnFreq} \Dissipator[\sqrt{2}\PositionOperator / \PosDim_{\UnFreq}]\DensityMatrix = - \DimNoise_{\UnFreq}  [\PositionOperator,[\PositionOperator,\DensityMatrix]]/\PosDim_{\UnFreq}^2$, where the diffusion strength $\DimNoise_{\UnFreq}$ can depend on the trapping frequency ${\UnFreq}$.
Within the short-time fidelity-decay estimate above, this dissipator gives
\begin{align}
    \DecoRate = 2 \DimNoise_{\UnFreq} \frac{(\Delta \PosDim)^2}{ \PosDim_{\UnFreq}^2}. 
\end{align}
To generate quantum non-Gaussian features from an initially Gaussian state, the non-Gaussian rate should dominate the decoherence rate,
\begin{align}
    \frac{\NonGaussianRate}{\DecoRate} = \sqrt{\frac{15}{4}}  \frac{|\NonLin_{\UnFreq}| \UnFreq}{\DimNoise_{\UnFreq} \PosDim_{\UnFreq}}  \Delta \PosDim = \sqrt{\frac{15}{4}} \frac{|\CubNonlinDim|}{\Mass \UnFreq \DimNoise_{\UnFreq} } \Delta \PosDim \gtrsim 1.
\end{align}
To date, the ground-state cooling of levitated particles has been achieved only via optical potentials, which typically yield $ \Delta \PosDim = \PosDim_{\Frequency}  \sim 10$ pm for the center-of-mass motion and $ \Delta \PosDim = \Delta \theta \sim 10$ {\textmu}rad for the libration.
These zero-point spreads lead to very small ratios $\NonGaussianRate / \GaussianRate$ and $\NonGaussianRate / \DecoRate$, so standard ground-state-engineering scenarios remain dominated by Gaussian dynamics.
Hence, wave-function delocalization, $\Delta \PosDim / \PosDim_{\Frequency} \gg 1$, has been proposed as a way to enhance nonlinear dynamics and generate squeezed cubic-phase~\cite{Rakhubovsky2021,Neumeier2024,Roda-Llordes2024b} or quartic-phase states~\cite{Rosiek2024}.
Here, we consider large wave-function expansions to access regimes where non-Gaussian dynamics is sufficiently coherent to enable, in principle, arbitrary state preparation within a finite Hilbert-space truncation.

However, in addition to the mechanical diffusion channel, strongly delocalized states can also become sensitive to higher-order noise mechanisms described by nonlinear collapse operators, for example, $\DimNoise^{(2)}_{\UnFreq}\Dissipator[\PositionOperator^2/\PosDim_{\UnFreq}^2]\DensityMatrix$~\cite{Riera-Campeny2024}.
Such channels naturally arise from fluctuations of the trapping curvature, including trap-frequency noise, optical intensity fluctuations, electrode-voltage noise, and more general multiplicative noise processes.
For Gaussian states with vanishing first moments, this dissipator induces a decoherence rate scaling as
\begin{align}
    \DecoRate^{(2)}= 2\DimNoise^{(2)}_{\UnFreq}\frac{(\Delta \PosDim)^4}{\PosDim_{\UnFreq}^4},
\end{align}
which grows faster with wave-function delocalization than mechanical diffusion.
Consequently, while motional expansion enhances the effective nonharmonic dynamics~\cite{Rosiek2024}, sufficiently strong higher-order noise can ultimately limit the achievable delocalization, purity, and maximum size of the prepared non-Gaussian states.
In particular, such nonlinear decoherence channels become increasingly relevant for states exhibiting large phase-space extent or fine interference structure, such as Schrödinger cat states, where they can rapidly suppress non-Gaussian coherence even in regimes where ordinary diffusion remains weak.
Motional delocalization should therefore be viewed as an intermediate control resource whose enhancement of coherent nonlinear dynamics must be balanced against the increased susceptibility to higher-order decoherence channels.

\section{The state preparation protocol}
\label{sec_state-prep}
\begin{figure*}[ht!]
    \includegraphics[width=\linewidth]{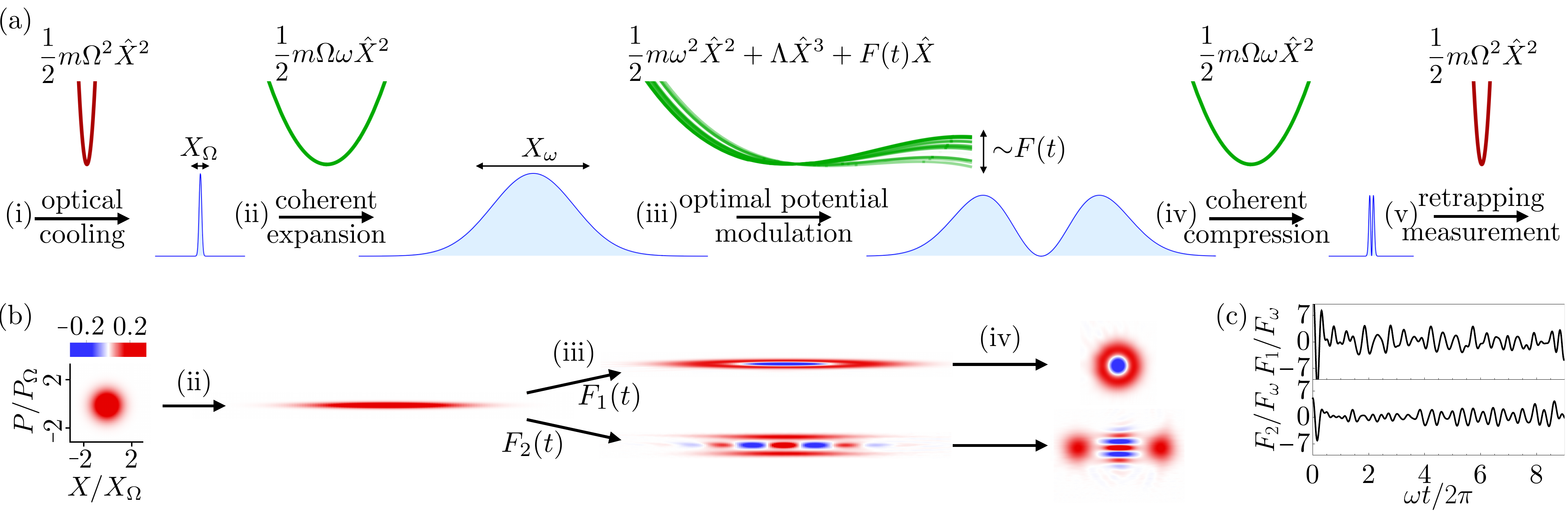}
    \caption{ (a) Sketch of the protocol for preparing non-Gaussian motional states of a massive particle by combining delocalization-induced enhancement of nonharmonicity with optimal potential modulation.
    First, the particle is cooled close to the ground state in a tightly confining optical potential (red). 
    It is then coherently expanded into the ground state of the shallow nonoptical potential.
    Optimal modulation of the nonharmonic potential then prepares the target non-Gaussian state, which is subsequently compressed back to the original trapping scale and measured in the optical potential.
    The strongly delocalized intermediate state enhances the weakly nonlinear dynamics during the control stage, while the final compression maps the generated state into a more tightly confined and better-protected regime compatible with optical readout.
    (b) Representative realizations of the protocol for preparing the first excited Fock state $\KetFockStateOne$ and an even Schrödinger cat state.
    In both cases, we set the nonharmonicity parameter $\NonLin_{\LowFrequency} = 0.04$, use a dimensionless control duration $\LowFrequency \DimTimeMax = 9 \cdot 2 \pi$, and choose a motional delocalization of $\MotionalDeloc = 14$ dB.
    The Wigner functions after the main stages of the protocol are shown.
    (c) Corresponding optimized linear control forces, $\DimForce_1(\DimTime)$ and $\DimForce_2(\DimTime)$, are shown. \label{Fig2}}
\end{figure*}

We now introduce a state-preparation protocol that combines delocalization-induced nonharmonicity enhancement with time-dependent modulation of the terms in~\eqref{potlandscape}.
The protocol is sketched in Fig.~\ref{Fig2}(a) and consists of five steps:
\begin{enumerate}[label=(\roman*)]
  \item Preparation of the motional ground state, or a near-ground-state thermal state, in a high-frequency optical potential;
  \item Coherent transfer to the ground state of a low-frequency, possibly nonoptical, potential; 
  \item Preparation of the delocalized target state via optimal control of the potential;
  \item Coherent compression of the state;
  \item Subsequent manipulation and measurement in either an optical or a nonoptical potential.
\end{enumerate}
We now describe these steps in detail.

\subsection{Preparation in an optical trap and transfer to the ground state of a shallow nonoptical potential}
Ground-state cooling has been achieved in optical tweezers~\cite{Delic2020,Magrini2021,Tebbenjohanns2021,Piotrowski2023} and optical lattices~\cite{Kamba2022,Kamba2023} using feedback-cooling schemes~\cite{Rossi2018}.
Here, we assume that the particle is initially prepared in a thermal state of an optical harmonic potential, $\Mass \Frequency^2 \PositionOperator^2/2$, with a low phonon occupation number $\AvN < 1$.
The next step transfers the state between the ground states of two harmonic potentials: a high-frequency optical potential, $\Mass \Frequency^2 \PositionOperator^2/2$, and a low-frequency, possibly nonoptical, potential, $\Mass \LowFrequency^2 \PositionOperator^2/2$, with $\LowFrequency \ll \Frequency$~\cite{Streltsov2021}.
Related protocols have been considered in the context of trap softening~\cite{Chen2010} and optimal control~\cite{Guery-Odelin2019}.
We use a phase-space rotation in an auxiliary harmonic potential whose frequency is the geometric mean of the initial and final frequencies~\cite{Chen2010},
\begin{align}
\HamiltonianExpand =  \frac{\MomentumOperator^2} { 2 \Mass} + \frac{1}{2} \Mass \Frequency\LowFrequency \PositionOperator^2,
\end{align} 
as a realization of such a state transfer.
Starting from the optical ground state, $\KetGroundState_{\Frequency}$, evolution under $\HamiltonianExpand$ for $\DimTimeExpand = \pi / (2 \sqrt{\Frequency \LowFrequency})$ prepares the ground state of the nonoptical potential with frequency $\LowFrequency$, $\KetGroundState_{\LowFrequency}$, yielding the motional delocalization, which we quantify in decibels as
\begin{align}
\MotionalDeloc [\rm{dB}] = 10 \log_{10}\rounds{\frac{\PosDim_{\LowFrequency}^2}{\PosDim_{\Frequency}^2}} = 10 \log_{10}\rounds{\frac{\Frequency}{  \LowFrequency}}.
\label{MotDec}
\end{align}
This expansion step can be further accelerated through optimal control~\cite{Guery-Odelin2019} or coherent inflation~\cite{Romero-Isart2017,Weiss2021}.
The purpose of this expansion is not to generate the final target state itself, but rather to transiently access a regime where weak intrinsic nonharmonicities become sufficiently strong to enable coherent non-Gaussian control.

\subsection{Potential modulation with static cubic nonharmonicity}
In step (iii), we use time-dependent control of the potential terms introduced in~\eqref{potlandscape}.
In a hybrid optical–nonoptical setup, such modulation may be implemented using appropriately designed electrodes whose applied voltages or currents are varied in time~\cite{Wilson2015,Foot2018,Magrini2021,Tebbenjohanns2020,Tebbenjohanns2021}.
For librational motion, an analogous control may be realized by temporally varying an additional circularly polarized optical field~\cite{Zielinska2023}.
The achievable nonharmonicity $\CubNonlinDim$, however, is constrained by the electrode geometry and other hardware limitations~\cite{Home2011,Rosiek2024}.
The effective nonharmonicity $\NonLin_{\LowFrequency}$, defined in Equation~\eqref{nonlinpar}, can be increased by reducing $\LowFrequency$, since $\NonLin_{\LowFrequency} \sim \LowFrequency^{-5/2}$.
This comes at the cost of a longer protocol, whose characteristic time scale grows as $\sim 1/\LowFrequency$.
Our control strategy keeps the quadratic and cubic terms fixed while optimally modulating the linear component of the potential,
\begin{align}
    \HamiltonianShake (\DimTime) = \hbar \LowFrequency \rounds{\frac{1}{2} \MomOp_{\LowFrequency}^2 + \frac{1}{2} \PosOp_{\LowFrequency}^2 +  \NonLin_{\LowFrequency} \PosOp_{\LowFrequency}^3 + \force(\DimTime)\PosOp_{\LowFrequency}  }.
    \label{potshake}
\end{align}
Here $\force(\DimTime) = \DimForce(\DimTime) \PosDim_{\LowFrequency} / \hbar \LowFrequency$ is the dimensionless control amplitude.
Equation~\eqref{potshake} is understood as a local effective Hamiltonian over the spatial region explored during the protocol.
Linear oscillator drives have been proposed and realized in systems with Kerr-type nonlinearities~\cite{Kamal2009,Yuan2025,Zhao2023} and are available in several spin–oscillator platforms~\cite{Heeres2017,Eickbusch2022,Matsos2024,Matsos2025}.
The system defined by Equation~\eqref{potshake} is approximately controllable, i.e., by modulating the linear term, one can realize any quantum state with the fidelity arbitrarily close to 1~\footnote{The controllability can be directly shown using modified~\cite{Zhao2023} Theorem 1 from~\cite{Merkel2009}.}.
To our knowledge, driven dynamics with a leading cubic nonlinearity has previously been reported only for superconducting nonlinear asymmetric inductive-element resonators~\cite{Eriksson2024} and trapped-ion systems with nonlinearity engineering~\cite{Bazavan2026}, although without pulse optimization.
Here, we apply quantum optimal control~\cite{Werschnik2007,Glaser2015,Koch2022} to the linear term.
Time-dependent potential engineering has been shown theoretically to enable high-fidelity preparation of diverse target states~\cite{Grochowski2025a} and has been demonstrated experimentally for several control tasks in atomic systems~\cite{Bucker2011,Bucker2013,vanFrank2016a,Schmidt2020,Dupont2021,Lam2021}.
To optimize the control $\force(\DimTime)$, we use the state-transfer fidelity with respect to the target state $\ket{\TargetState}_{\LowFrequency}$ as the reward function.
After evolving $\KetGroundState_{\LowFrequency}$ under Equation~\eqref{potshake} for a duration $\DimTimeMax$, this fidelity is
\begin{align}
\Fidelity = \left|\braket{\WaveFunction(\DimTime = \DimTimeMax)}{\TargetState}_{\LowFrequency} \right|^2.
\end{align}
The subscript ${}_\LowFrequency$ indicates that the target is expressed in the oscillator basis of the low-frequency potential.
Relative to the final target $\ket{\TargetState}_{\Frequency}$, it is transformed by the squeezing and phase-space rotation associated with the subsequent compression step.
For the representation and optimization of the control function, we utilize the dressed chopped random-basis technique~\cite{Muller2022} with the gradient-free Nelder-Mead  algorithm~\cite{Nelder1965}.
In this method, the control is expanded in a finite set of sinusoidal functions with randomized frequencies, subject to a high-frequency cutoff chosen to represent an experimentally accessible bandwidth.
The optimization is performed using the open-source library QuOCS~\cite{Rossignolo2023}.

\subsection{Compression, manipulation, and readout}
While strong motional delocalization enhances weak nonharmonic effects~\cite{Rosiek2024} and enables efficient quantum control, highly expanded states are generally more sensitive to decoherence, technical noise, and measurement imperfections.
The compression step, therefore, coherently maps the generated non-Gaussian state back to the original tightly confining optical trap, where its spatial extent and susceptibility to position-dependent noise are reduced and it becomes compatible with optical manipulation and readout.
The ultimate objective of the protocol is to prepare target non-Gaussian states in the original trapping configuration rather than leave them in the transiently delocalized regime.
The compression step uses the same intermediate harmonic evolution as the expansion step, producing the inverse rescaling together with an additional phase-space rotation~\cite{Weiss2021,Roda-Llordes2024b,Ji2026}.
The expansion and compression sequence also introduces a $\pi / 2$ phase-space rotation.
This rotation is irrelevant for the initially rotationally symmetric state used before the control step, but it must be included when defining the non-Gaussian target state,
\begin{align}
    \ket{\TargetState}_{\LowFrequency} = \exp\Big(\ImagUnit \pi \Creation_{\sqrt{\Frequency\LowFrequency}} \Annihilation_{\sqrt{\Frequency\LowFrequency}}/2\Big)\SqueezingOperator(\Squeeze)\ket{\TargetState}_{\Frequency}.
    \label{final_state_map}
\end{align}
Equation~\eqref{final_state_map} is obtained by applying the inverse compression map to the desired final state.
Because the squeezing and rotation do not commute, their ordering and the sign of the rotation follow from the explicit quarter-period evolution.
Here, $\SqueezingOperator(\Squeeze) =  \exp[\Squeeze(\Annihilation_{\Frequency}^{2}-\Annihilation_{\Frequency}^{\dagger 2})/2]$ is the squeezing operator with $\Squeeze =[ \log (\Frequency / \LowFrequency)]/4$ and $\ket{\TargetState}_{\Frequency}$ is the target state at the end of the whole protocol. 

After the compression step, the state needs to be measured and certified.
This may be achieved via several strategies, including optical readout or coupling to auxiliary two-level systems while the particle remains in a nonoptical potential~\cite{Streltsov2021}.
Optical readout may involve reconstruction of spatial probability distributions~\cite{Neumeier2024,Riera-Campeny2026}, potentially enhanced through quantum retrodiction~\cite{Rossi2019,Barnett2021,Lammers2024} or machine-learning methods~\cite{Weiss2019}.
Full state tomography may instead be obtained from a sequence of quadrature measurements, for example using time-of-flight expansion~\cite{Brown2023,Kamba2023}.
Access to fine phase-space structure may additionally be improved by coherent inflation in an inverted potential before measurement~\cite{Romero-Isart2017,Pino2018,Neumeier2024,Roda-Llordes2024b}.

\subsection{Time scales and protocol duration}
The protocol contains three dynamical steps: expansion of duration $\DimTimeExpand$, optimal control of duration $\DimTimeMax$, and compression of duration $\DimTimeExpand$.
The total duration of the protocol is therefore
\begin{align}
    \Frequency \DimTime_{\ProtocolLetter} = 2 \Frequency \DimTimeExpand + \Frequency \DimTimeMax = \pi 10^{\MotionalDeloc/20} + \LowFrequency \DimTimeMax 10^{\MotionalDeloc/10}.
\end{align}
Throughout the analyzed protocols, we typically use $\LowFrequency \DimTimeMax \sim 10 \pi$.
For the parameters considered here, the control step is substantially longer than the combined expansion and compression steps,
\begin{align}
    \frac{\DimTimeMax}{2 \DimTimeExpand} = \frac{\LowFrequency \DimTimeMax}{\pi} 10^{\MotionalDeloc/20},
\end{align}
which can be of order of $100$ for parameters we consider across the manuscript.
This reflects the central tradeoff of the protocol: decreasing $\LowFrequency$ enhances the dimensionless nonharmonicity and facilitates coherent non-Gaussian control, but also increases the physical duration of the protocol and hence its exposure to decoherence.

\section{Non-Gaussian state preparation}
\label{sec_nongauss}
In this section, we analyze several applications of the protocol introduced in Section~\ref{sec_state-prep}.
As our primary example, we consider preparation of the first excited Fock state and analyze its scaling with the nonharmonicity and control duration.
To investigate the preparation of more complex states relevant to metrology and quantum technologies, we also consider high-number Fock states and large-amplitude Schr\"odinger cat states.

\subsection{First excited Fock state}
\begin{figure}[t!]
    \includegraphics[width=\linewidth]{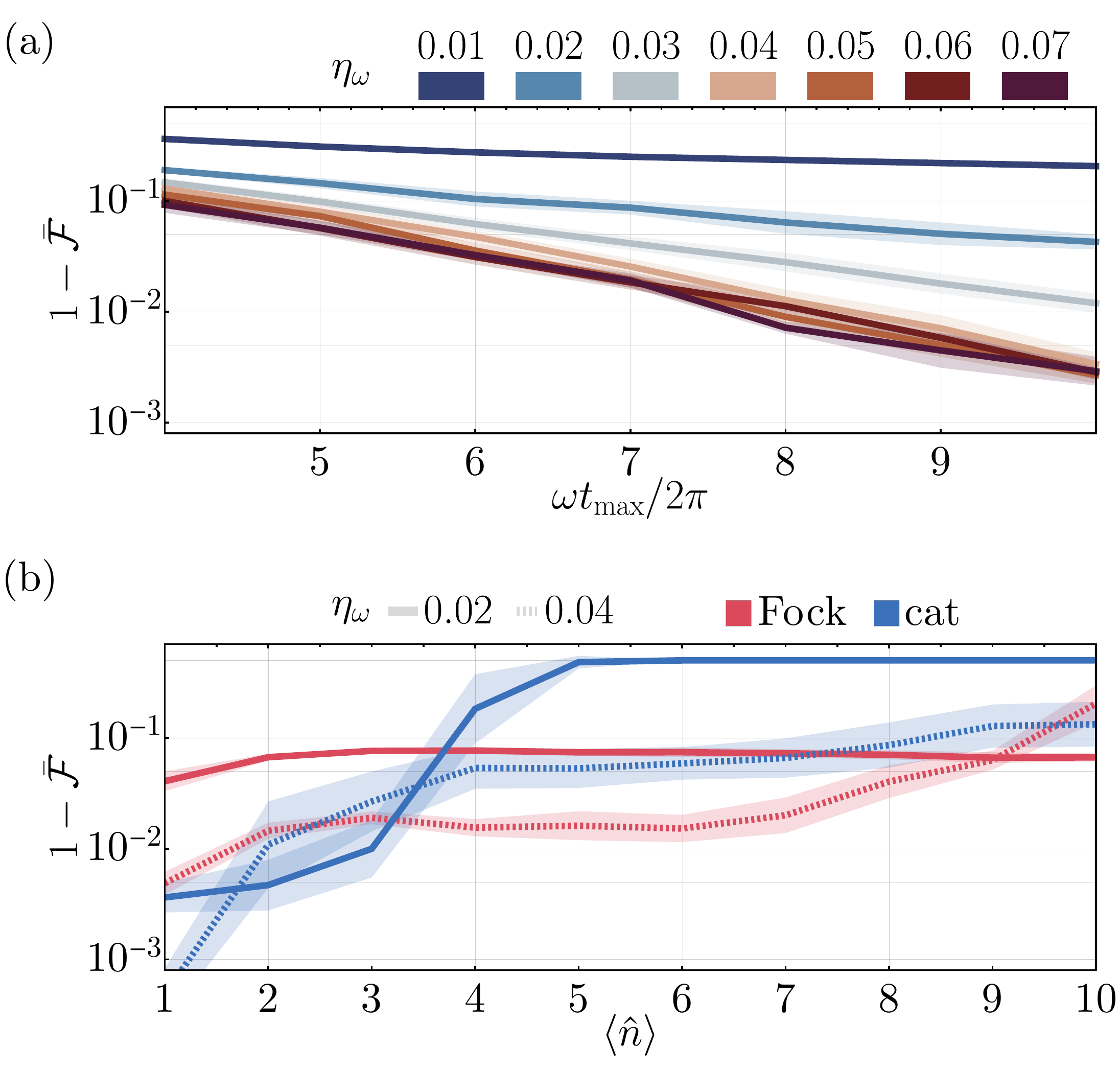}
    \caption{Results for the quantum non-Gaussian state preparation with an optimally controlled weakly cubic potential. (a) Geometric-mean infidelity for preparation of the first excited Fock state $\KetFockStateOne$, shown as a function of the control duration $\DimTimeMax$ and nonharmonicity $\NonLin_{\LowFrequency}$. The shaded area corresponds to half of the standard deviation of the log-normal distribution in the set of optimization runs. Increasing $\NonLin_{\LowFrequency}$ beyond approximately 0.04 yields little additional reduction in infidelity. (b) Geometric-mean preparation infidelity for Fock states $\KetFockStateN$ and even cat states as a function of their target mean occupation $\left<\hat{n}\right>$ and varying nonharmonicity $\NonLin_{\LowFrequency}$. \label{Fig3}}
\end{figure}

We first consider preparation of the single-phonon Fock state, $\KetFockStateOne$.
Selective preparation of a single excitation can be achieved through coupling to an auxiliary two-level system~\cite{Law1996,Bergholm2019,Podhora2022,Yang2024} or through a beam-splitter interaction with another bosonic mode initially prepared in a Fock state~\cite{Romero-Isart2011c,Gao2019}.
In weakly nonharmonic bosonic systems, by contrast, the small anharmonic splitting makes individual transitions difficult to resolve.
Preparation of $\KetFockStateOne$ therefore generally requires either weak, spectrally selective driving~\cite{Serwane2011,Serwane2011a} or quantum optimal control~\cite{vanFrank2016a,Grochowski2025a}.
In Fig.~\ref{Fig2}(b), we show an example of such a state preparation protocol involving a single best optimization run for a given $\NonLin_{\LowFrequency} = 0.04$ and $\LowFrequency \DimTimeMax = 9 \cdot 2 \pi$ that yields $1 - \Fidelity \approx 0.005$.
For each set of physical parameters, many optimization runs were performed, each with a different initial control pulse and with a fixed computational time.
As our figure of merit, we choose to take the geometric mean of the final infidelities, $1-\AvFidelity $ (see Appendix~\ref{appA} for details),
\begin{align}
    1-\AvFidelity  = \exp\squares{\frac{1}{\NoRuns} \sum_{i=1}^{\NoRuns} \log (1- \Fidelity_i)}.
    \label{avfif}
\end{align}
Here, $i$ labels the $\NoRuns$ independent optimization runs.
The observed infidelity distributions are approximately log-normal over the parameter range considered.
A representative histogram is shown in Appendix~\ref{appA}.
Quantity~\eqref{avfif} equals the sample geometric mean of the infidelity and, for a log-normal distribution, estimates its median.
This statistic characterizes the typical performance of the chosen optimization procedure for a fixed parameter set while reducing sensitivity to unusually good or poor runs.
In Fig.~\ref{Fig3}(a), we plot $1-\AvFidelity $ as a function of $\DimTimeMax$ and $\NonLin_{\LowFrequency}$.
Over the range of control durations studied, the approximate scaling we obtain is exponential, $\log(1-\AvFidelity) \propto -\PowScal \DimTimeMax$, where $\PowScal$ initially increases with $\NonLin_{\LowFrequency}$ to saturate for $\NonLin_{\LowFrequency} \gtrsim 0.04$.
This saturation suggests that a further increase in nonharmonicity does not improve the ability to prepare the state.
The relevant time scale for this part of the protocol is given by $1/\LowFrequency$, outperforming the time scale implied by the uneven splitting of energy levels in nonharmonic potential, $1/\LowFrequency \NonLin_{\LowFrequency}^2$~\footnote{For cubic perturbation, term linear in $\NonLin_{\LowFrequency}$ vanishes.}, that enables a weak drive from the ground to the first excited state.
Unlike the behavior reported for some related control problems~\cite{Caneva2009,Rojan2014,vanFrank2016a}, our optimized infidelities do not display a clear cosine-squared dependence on the control duration.
Such behavior has previously been used to extract an effective quantum-speed-limited minimum transfer time~\cite{Bhattacharyya1983,Pfeifer1993,Caneva2009}.
A cosine-squared envelope may nevertheless emerge over a broader range of control durations or under a more exhaustive optimization.

\subsection{High-number Fock states and Schr\"odinger cat states}
As further examples, we consider high-number Fock states $\KetFockStateN$ and symmetric cat states~\cite{Rojan2014}, $\KetCatState_{\Frequency} = \Normalization (\ket{\CohDisLoop}_{\Frequency} + \ket{-\CohDisLoop}_{\Frequency})$, where $\ket{\CohDisLoop}_{\Frequency}$ is a coherent state, $\Normalization$ is a normalization constant, and we consider a spatial superposition, implying a real $\CohDisLoop$.
Note that, due to additional phase-space rotation, the latter task involves a state transfer to the momentum superposition of coherent states, $\ket{\TargetState}_{\LowFrequency} = \Normalization (\ket{\ImagUnit\CohDisLoop}_{\LowFrequency} + \ket{-\ImagUnit\CohDisLoop}_{\LowFrequency})$.

In Fig.~\ref{Fig3}(b), we plot the geometric-mean infidelity for the two state families as a function of the target mean phonon occupation, up to $\left<\hat{n}\right> = 10$.
For an even cat state with real $\CohDisLoop$, $\left<\hat{n}\right> = \CohDisLoop^2 \tanh \CohDisLoop^2$.
We fix the control duration $\LowFrequency\DimTimeMax = 9 \cdot 2 \pi$ and consider two values of $\NonLin_{\LowFrequency}$.
For low-number Fock states, the larger value of $\NonLin_{\LowFrequency}$ yields lower infidelity. For higher-number states, however, this trend reverses.
We attribute this reversal to the increasing spatial extent of high-number Fock states.
These states probe regions where the cubic term is stronger and where the locally cubic potential becomes increasingly asymmetric and eventually unstable on one side.
This both complicates control and reduces the validity of the truncated potential model.
This behavior also indicates that higher-order terms may be required both to preserve the validity of the potential model over the explored spatial range~\cite{Roda-Llordes2024b} and to stabilize the dynamics~\cite{vanFrank2016a}.
For cat-state preparation, the achieved infidelity likewise increases with the target occupation.
Larger cats require the creation of more widely separated branches and finer interference structure, making them more demanding under the fixed control duration and optimization budget~\cite{Rojan2014}.
For the smaller nonharmonicity $\NonLin_{\LowFrequency}$, the optimizer does not find high-fidelity large-cat solutions within the chosen computational budget.
Instead, all runs converge to states close to a single coherent state.
This behavior may be improved by reward functions tailored to cat-state structure~\cite{Krauss2023} or by alternative optimization methods.

\section{Experimental feasibility}
\label{sec_exp}
The protocol requires two main ingredients---sufficiently strong effective nonharmonicity $\NonLin_{\LowFrequency}$ and sufficiently long coherence times.
For center-of-mass motion, representative cubic coefficients in ion-trap geometries are of order $\CubNonlinDim_\text{ref} \sim 10^{-5}$ N/m$^2$~\cite{Home2011}.
Values up to approximately one order of magnitude larger may be achievable by positioning the particle closer to electrode surfaces, although this would generally increase technical noise and surface-induced decoherence.
Consider a charged silica nanoparticle of diameter 80 nm, initially cooled in an optical harmonic trap with $\Frequency = 2\pi \cdot 80$ kHz.
We choose the shallow-trap parameters such that $\NonLin_{\LowFrequency} = 0.04$, the value beyond which no further improvement was observed in the numerical optimizations of Sec.~\ref{sec_state-prep}.
Using Equation~\eqref{nonlinpar}, the required delocalization is estimated to scale as
\begin{align}
    \MotionalDeloc \approx 30 - 4 \log_{10} |\CubNonlinDim| / \CubNonlinDim_\text{ref} \ \ [\text{dB}].
\end{align}
For $\CubNonlinDim = \CubNonlinDim_\text{ref}$, the estimate gives $\LowFrequency/2\pi \sim 0.1$ kHz, corresponding to the motional delocalization of $\MotionalDeloc \sim 30 $ dB.
The characteristic dynamical time is therefore $1/\LowFrequency \sim 1.6$ ms.
For librational motion, the optical potential can be written in the form
\begin{align}
    \frac{\Pot_{\text{l}}  (\LibAngle)}{ \hbar \Frequency_{\text{l}}} = \frac{1}{2 \LibNonLin^2}\sin^2(\LibNonLin \LibAngle),
\end{align}
where $\LibNonLin = \sqrt{\hbar / \InertiaMoment \Frequency_{\text{l}}} \approx 10$ {\textmu}rad~\cite{Dania2025,Troyer2026} is the angular oscillator length, $\InertiaMoment $ is the moment of inertia, $\Frequency_{\text{l}}$ is the libration frequency, and $\LibAngle$ is the dimensionless angular coordinate expressed in oscillator units.
Expanding the potential around an operating point $\LibAngle_0$, the local cubic nonharmonicity is
\begin{align}
    \NonLin_\LowFrequency = -\frac{\LibNonLin \sin(2 \LibNonLin \LibAngle_0)}{3 \cos^{5/4}(2 \LibNonLin \LibAngle_0)}  ,
\end{align}
while the corresponding local harmonic frequency is
\begin{align}
    \LowFrequency = \sqrt{\cos(2 \LibNonLin \LibAngle_0)} \Frequency_{\text{l}}.
\end{align}
For the libration frequency $\Frequency_{\text{l}} / 2 \pi = 1$ MHz and moment of inertia $\InertiaMoment = 1.7 \cdot 10^{-31}$ kg$\cdot$m$^2$, the saturation nonharmonicity $\NonLin_\LowFrequency = 0.04$ then corresponds to $\LowFrequency \approx 0.02 \Frequency_{\text{l}} = 2 \pi \cdot  0.02$ MHz and expansion of 16 dB.
The strongly delocalized configuration is used only during the intermediate expansion and control steps.
After compression, the prepared state is returned to the tighter trap.
This reduces its exposure to position-dependent decoherence when largely delocalized.
    
We next consider the principal sources of state-preparation error in a levitated platform:
\begin{enumerate}[label=(\textsc{\roman*})]
\item residual initial thermal occupation $\AvN$,
\item background-gas collisions,
\item thermal emission from the levitated object~\cite{Romero-Isart2011a,Hackermuller2004a,Bateman2014,Agrenius2023},
\item vibrations, stray forces, and technical field noise~\cite{Gehm1998,Henkel1999,Schneider1999,Brownnutt2015,Kumph2016},
\item control imperfections~\cite{Grochowski2025a},
\item static parameter and calibration uncertainties.
\end{enumerate}
Residual thermal occupation limits the fidelity attainable by unitary control, since the spectrum of the initial density operator is preserved.
For an initial thermal state, the maximum overlap with any pure target is bounded by its largest eigenvalue, $1/(1+\AvN)$. 
This limitation can be reduced through improved cooling~\cite{Popp2006,Grochowski2025a}.
Background-gas collisions can be strongly suppressed by operating in ultrahigh vacuum~\cite{Dania2024}.
We model the stochastic contributions from sources (\textsc{iii})-(\textsc{v}) using a mechanical-diffusion master equation, while static parameter uncertainties in source (\textsc{vi}) are included through an additional Hamiltonian correction,
\begin{align}
    \partial_{\DimTime} \DensityMatrix= \frac{1}{i \hbar} \squares{\Hamiltonian_{\ExpandLetter / \ShakeLetter} (\DimTime)+\HamiltonianCorr,\DensityMatrix} - \frac{ \DimNoise_{\ExpandLetter / \ShakeLetter}}{\PosDim_{\Frequency}^2} \squares{\PositionOperator,\squares{\PositionOperator,\DensityMatrix}}.
    \label{mastereq}
\end{align}
Here, $\DensityMatrix$ describes the motional state.
The double-commutator term represents momentum diffusion for center-of-mass motion, or angular-momentum diffusion for libration~\cite{Romero-Isart2011}.
Under the approximations used here, thermal-radiation recoil, technical force noise, and control noise are absorbed into the effective rates $\DimNoise_{\ExpandLetter / \ShakeLetter}$~\cite{Roda-Llordes2024b,Riera-Campeny2024,Dania2025}.
We normalize the diffusion operator by the optical oscillator length $\PosDim_{\Frequency}$.
Consequently, the numerical values of $\DimNoise_{\ExpandLetter}$ and $\DimNoise_{\ShakeLetter}$ refer to this convention.
The subscripts e and s label the expansion/compression steps and the state-preparation control step, respectively.
Static calibration and parameter errors are modeled through the Hamiltonian correction,
\begin{equation}
    \HamiltonianCorr = \corrForce \PositionOperator + \frac{1}{2} \Mass \corrFreq^2 \PositionOperator^2 + \corrNonlin \PositionOperator^3.
\end{equation}
Here, $\corrForce$, $\corrFreq$, and $\corrNonlin$ denote static errors in the linear-force coefficient, harmonic curvature, and cubic coefficient, respectively.
We solve~\eqref{mastereq} in the weak-error regime, where the decrease of the fidelity~\cite{Uhlmann1976,Jozsa1994,Liang2019},
\begin{align}
\Fidelity \rounds{\DensityMatrix,\DensityMatrix_0} = \rounds{\text{tr} \sqrt{\sqrt{\DensityMatrix} \DensityMatrix_0 \sqrt{\DensityMatrix}}}^2,
\end{align}
from 1 is small.
Here, $\DensityMatrix_0$ denotes the ideal state obtained from noiseless evolution with the nominal Hamiltonian.
For the expansion and compression steps, we evaluate the diffusion-induced fidelity loss analytically in Appendix~\ref{appB} and the effect of static Hamiltonian errors perturbatively in Appendix~\ref{appC}.
For the optimal-control step, the corresponding errors are evaluated numerically in Appendix~\ref{appE}.
The resulting parameter scalings for the different noise and calibration errors are summarized in these appendices.

\begin{figure}[t!]
    \includegraphics[width= \linewidth]{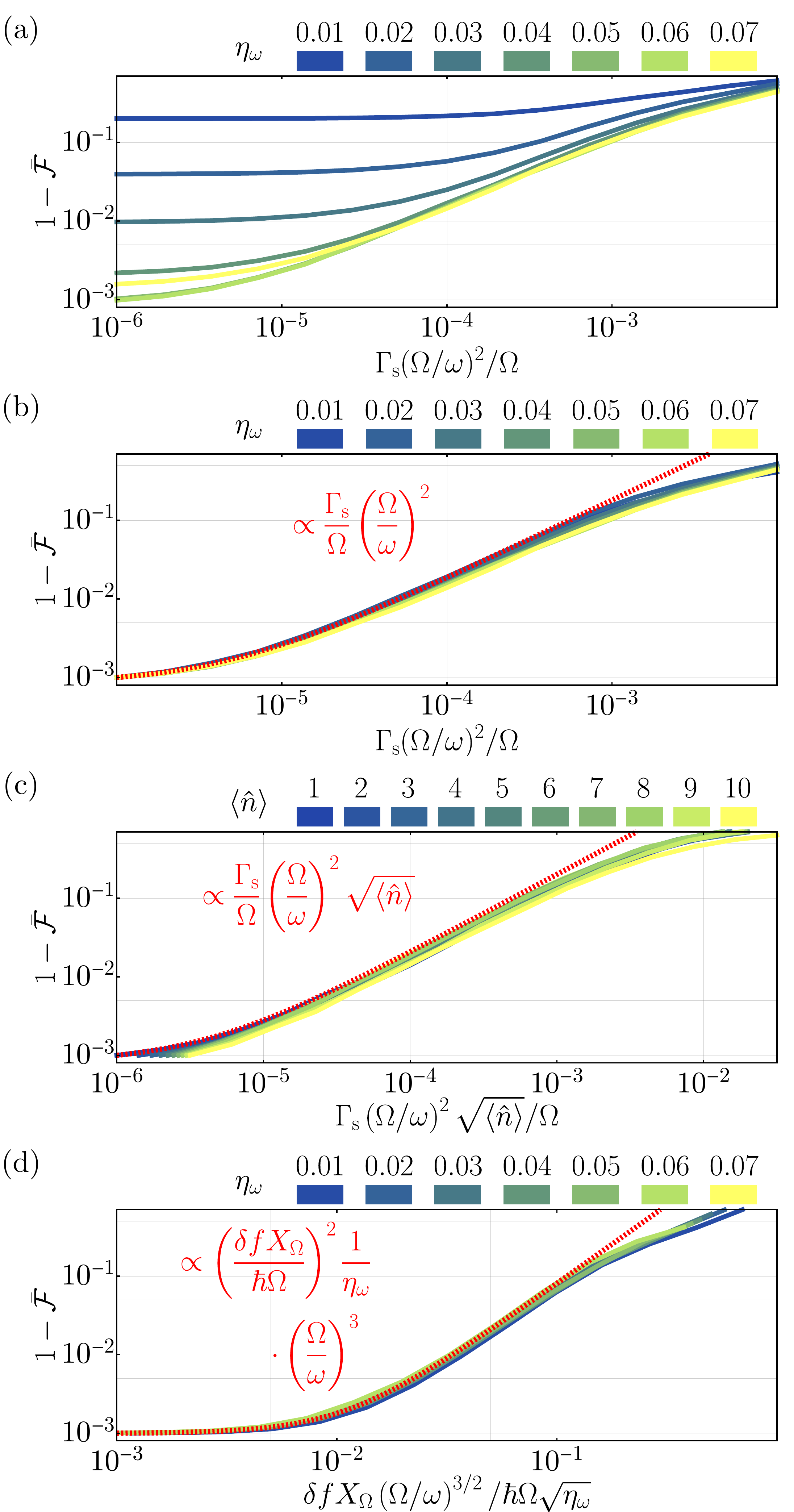}
    \caption{Effects of diffusion and static control errors on the optimal-control stage of the state-preparation protocol.
    (a) Minimum geometric-mean infidelity for preparation of the first excited Fock state as a function of the control-stage diffusion rate $\DimNoise_{\ShakeLetter}$, for several values of $\NonLin_{\LowFrequency}$.
    At each $(\DimNoise_{\ShakeLetter},\NonLin_{\LowFrequency})$, the result is minimized over the considered control durations $\DimTimeMax$.
    (b) Excess infidelity after subtraction of the zero-noise baseline, $1-\AvFidelity \rightarrow \AvFidelity (\DimNoise_\ShakeLetter = 0) -\AvFidelity $.
    An approximate common scaling with $\DimNoise_{\ShakeLetter}$ is revealed. 
    (c) Corresponding excess infidelity for preparation of higher-number Fock states. The data exhibit the additional approximate dependence $\DecoCoeffMain_\ShakeLetter \propto \left<\hat{n}\right>^{1/2}$.
    (d) Infidelity induced by a static force offset in the control Hamiltonian.
    \label{Fig4}}
\end{figure}

More specifically, Fig.~\ref{Fig4}(a) shows, for preparation of $\KetFockStateOne$, the minimum geometric-mean infidelity at fixed $\DimNoise_\ShakeLetter$ and $\NonLin_\LowFrequency$, optimized over the considered control durations $\DimTimeMax$.
At each $\DimTimeMax$, every optimized pulse is reevaluated in the presence of diffusion, and the corresponding infidelities are geometrically averaged over the independent optimization runs.
For smaller $\NonLin_\LowFrequency$ , the baseline coherent-control infidelity is larger [cf. Fig.~\ref{Fig3}(a)], so the additional noise-induced degradation becomes visible only at larger $\DimNoise_\ShakeLetter$.
This does not by itself imply greater intrinsic robustness.
Subtracting the zero-noise baseline, $1-\AvFidelity \rightarrow \AvFidelity (\DimNoise_\ShakeLetter = 0) -\AvFidelity $, causes the curves to approximately collapse onto a common noise-induced contribution.
Within the range studied, the excess infidelity is well approximated by $1-\AvFidelity = \DecoCoeffMain_\ShakeLetter \DimNoise_\ShakeLetter (\Frequency/ \LowFrequency)^{2} / \Frequency $ with a nearly $\NonLin_\LowFrequency$-independent coefficient $\DecoCoeffMain_\ShakeLetter$.
For the expansion and compression steps, the corresponding diffusion-induced infidelity scales as $1-\Fidelity = \DecoCoeffMain_\ExpandLetter \DimNoise_\ExpandLetter (\Frequency/ \LowFrequency)^{3/2} / \Frequency $.
The stronger exponent in the control-stage scaling indicates a more rapid increase with motional delocalization, although the absolute errors also depend on the coefficients $\DecoCoeffMain_\ShakeLetter$ and $\DecoCoeffMain_\ExpandLetter$.
The same delocalization exponents are observed for higher target occupations, but the prefactors depend on the target state.
For the cases studied, the expansion/compression coefficient scales approximately as $\DecoCoeffMain_\ExpandLetter \propto \left<\hat{n}\right>$, whereas the control-stage coefficient follows the weaker empirical dependence $\DecoCoeffMain_\ShakeLetter \propto \left<\hat{n}\right>^{1/2}$ [cf. Appendix~\ref{appB} and Fig.~\ref{Fig4}(c), respectively].
Thus, although the control stage is more sensitive to delocalization, its noise-induced error grows more slowly with target occupation.
We obtain analogous power-law dependences for static parameter errors.
Fig.~\ref{Fig4}(d), for example, shows the infidelity induced by a constant force offset during the control stage.
The expansion/compression and optimal-control steps again exhibit different dependences on the delocalization.
Together, these results provide estimates of the coherence requirements and parameter-control precision needed to implement the protocol.

\section{Beyond single-particle state preparation}
\label{sec_beyond}
So far, we have shown that potential modulation can be used for state preparation in a weakly nonharmonic system.
Related forms of time-dependent potential control have been developed in systems with much stronger nonlinearities, including state preparation in Bose–Einstein condensates~\cite{Hohenester2007,Bucker2011,Bucker2013a,Jager2014,vanFrank2016,Hocker2016,Sorensen2018,Schmidt2020,Dupont2021,Xu2022b}, fast atom transport~\cite{Calarco2004,Dorner2005,Lam2021}, optimal control of single neutral atoms~\cite{Kendell2024,Grochowski2025a}, ion shuttling~\cite{Sterk2022}, cat-state generation through two-photon driving~\cite{Puri2017}, lattice interferometry~\cite{Weidner2018}, and emulation of spin–boson dynamics~\cite{Hwang2025}.
Beyond state preparation, time-dependent potential control has also been proposed for implementing unitary operations, state discrimination, squeezing, and cooling in single-particle systems~\cite{Grochowski2025a}.
Here, we go beyond these efforts by preparing mechanical Bell states in two-mode systems, considering both two interacting particles and two motional modes of a single particle.
In both cases, the dominant intermode dynamics is Gaussian, while weak local or cross corrections provide the required nonlinearity.
This provides an alternative to protocols generating purely Gaussian interparticle entanglement~\cite{Krisnanda2020,Rakhubovsky2020,Rudolph2020,Qvarfort2020,Weiss2021,Cosco2021,Brandao2021,Chauhan2022,Rudolph2022} and to schemes in which non-Gaussian entanglement is mediated by auxiliary qubits~\cite{Martinetz2020}.
Unlike two-mode squeezed Gaussian states, the Bell states considered here are intrinsically non-Gaussian and can support forms of nonclassical correlations relevant to more stringent tests of quantum mechanics~\cite{Walschaers2021}.

\begin{figure}[t!]
\includegraphics[width=\linewidth]{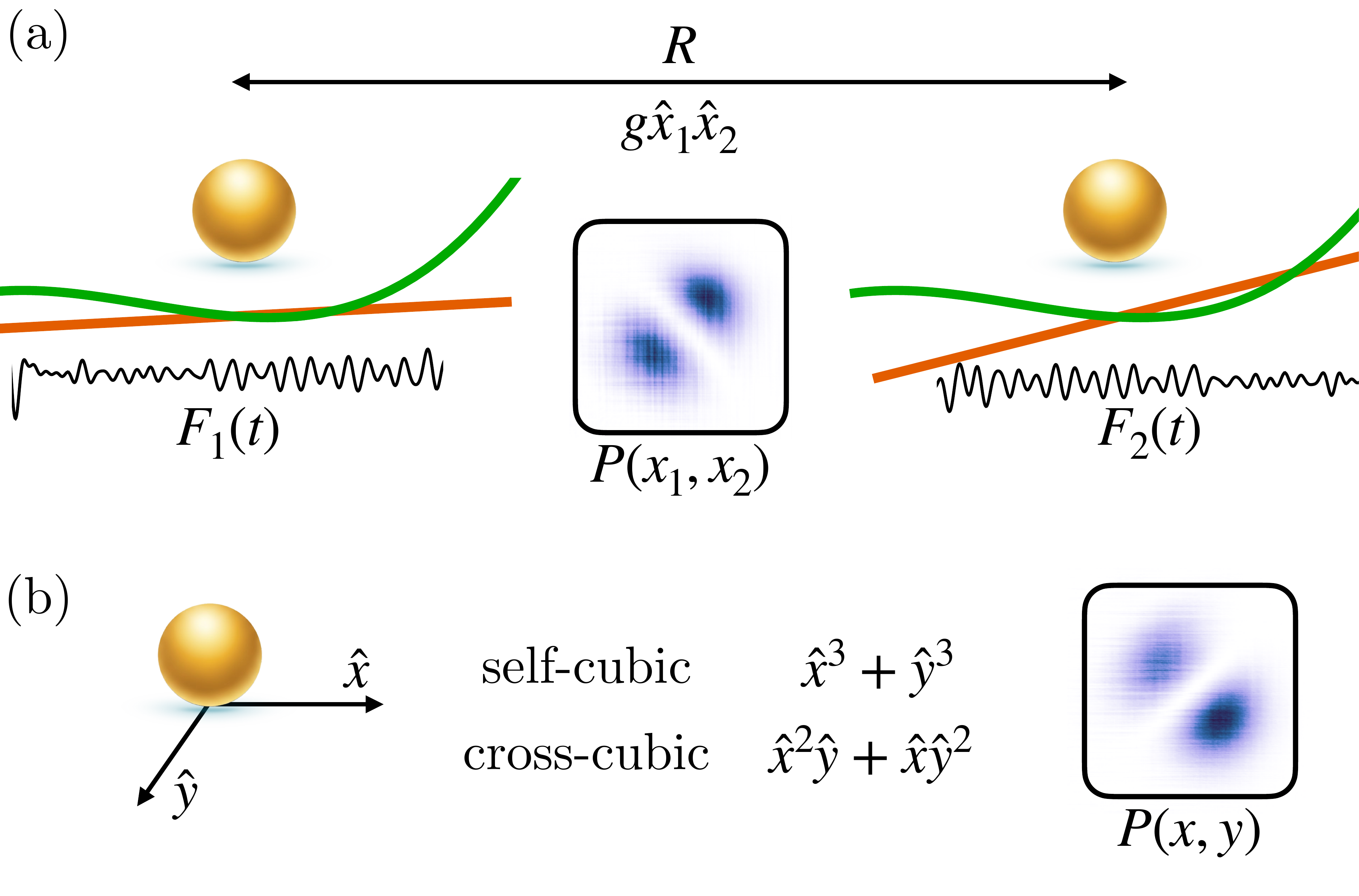}
    \caption{Configurations considered for two-mode Bell-state generation.
    (a) Two particles separated by a distance $\SepD$, each confined in a local weakly nonharmonic potential, interact through a long-range force.
    Motional delocalization increases the dimensionless interaction strength during the control stage.
    The interparticle coupling is bilinear and therefore Gaussian, while the required nonlinearity is provided by the local cubic terms.
    Independent modulation of the two local potentials then enables preparation of the symmetric mechanical Bell state.
    The inset shows the optimized joint position probability density, $P(\Pos_1,\Pos_2)$.
    (b) A single particle is simultaneously delocalized in two motional modes.
    In addition to the self-cubic terms of the individual modes, the nonseparable potential generates cross-cubic terms that couple them.
    This weak nonlinear coupling enables preparation of an antisymmetric Bell state.
    The inset shows the corresponding optimized joint position probability density, $P(\Pos,\PosY)$.\label{Fig5}}
\end{figure}

\subsection{Two interacting particles}
The first protocol involves two particles separated by a distance $\SepD$ and interacting with each other (see Fig.~\ref{Fig5}).
For simplicity, we assume that the particles are identical, i.e., they have the same mass $\Mass$ and their interaction is fully reciprocal.
Each particle experiences the same local potential and undergoes the expansion protocol of Sec.~\ref{sec_state-prep}, transferring from the ground state of a tightly confining optical trap of frequency $\Frequency$ to that of a shallow nonoptical trap of frequency $\LowFrequency$.
They also experience the same cubic nonharmonicity $\CubNonlinDim$ in the shallow nonoptical potential.
This assumption simplifies the theoretical model but does not alter the qualitative physical picture of two similar but not identical particles, whose dynamics can be solved numerically.
The particles interact via a central potential,
\begin{align}
    \Pot(\PositionOperator_1 - \PositionOperator_2) = \frac{\Pot_0}{\left[\SepD^2 + (\PositionOperator_1 - \PositionOperator_2)^2 \right]^{\alpha/2}}.
    \label{cen_pot}
\end{align}
For $\alpha = 1$ and $\Pot_0 = \Charge_1 \Charge_2 / 4 \pi \varepsilon_0$, Equation~\eqref{cen_pot} describes the Coulomb interaction between particles whose equilibrium separation $\SepD$ is orthogonal to the displacement direction $\PositionOperator$.
Here, $\Charge_1$ and $\Charge_2$ are the particle charges and $\varepsilon_0$ is the vacuum permittivity.
The same form also captures the leading distance dependence of other central interactions, including gravity for $\alpha = 1$, while suitable effective exponents may approximate dispersion forces in particular geometries~\cite{Weiss2021}.
Expanding the interaction to second order in $|\PositionOperator_1 - \PositionOperator_2|/\SepD$, the Hamiltonian can be written as 
\begin{align}
    \frac{\Hamiltonian}{\hbar \UnFreqT} = \sum_{i=1}^2 \left( \frac{1}{2} \MomOp_{i,\UnFreqT}^2 + \frac{1}{2} \PosOp_{i,\UnFreqT}^2 \right) + \frac{\IntG_\UnFreq} {2-\IntG_\UnFreq} \PosOp_{1,\UnFreqT}\PosOp_{2,\UnFreqT},
    \label{linearized}
\end{align}
where $\UnFreqT^2 = \UnFreq^2(1-\IntG_\UnFreq/2)$, $\UnFreq$ is the bare trapping frequency, and the dimensionless coupling constant is
\begin{align}
\IntG_\UnFreq = 2\alpha \Pot_0 / \Mass \UnFreq^2 \SepD^{\alpha+2}.
\end{align}
The ground state $\TwoModeGroundState_{\UnFreq,\IntG_\UnFreq}$ of Hamiltonian~\eqref{linearized} is separable in the center-of-mass and relative normal modes but entangled in the local particle basis~\cite{Hong-Yi1992,Zhou2020},
\begin{align}
\TwoModeGroundState_{\UnFreq,\IntG_\UnFreq} = \KetGroundState_{+,\UnFreq} \KetGroundState_{-,\UnFreq_-}.
\end{align}
Here, $\KetGroundState_{+,\UnFreq}$ is the ground state of the harmonic oscillator with frequency $\UnFreq$ that describes the center-of-mass motion of two particles, $\PositionOperator_+ = (\PositionOperator_1 + \PositionOperator_2) /\sqrt{2}$.
Similarly, $\KetGroundState_{-,\UnFreq_-}$ is the ground state of the relative motion, $\PositionOperator_- = (\PositionOperator_1 - \PositionOperator_2) /\sqrt{2}$, which is given by the harmonic oscillator with frequency $\UnFreq_- = \UnFreq \sqrt{1 - \IntG_\UnFreq}$.
The stability condition for this Hamiltonian is given by $ \IntG_\UnFreq < 1$.
Protocols exploiting the unstable regime have also been considered~\cite{Weiss2021}.
We assume that the interacting ground state $\TwoModeGroundState_{\Frequency,\IntG_\Frequency}$ is prepared in the tightly confining optical trap, for example through simultaneous feedback cooling of the center-of-mass and relative modes~\cite{Rudolph2022}.
In the regime $\IntG_\Frequency \ll 1 $, this state is close to the product of the local optical ground states.
We consider unaltered version of the expansion step, where each of the particles evolve in the potential $\HamiltonianExpand =  \MomentumOperator^2 / 2 \Mass +  \Mass \Frequency\LowFrequency \PositionOperator^2/2$ for $\DimTimeExpand = \pi / 2 \sqrt{\Frequency \LowFrequency}$, however with additional interaction between them given by $\IntG_{\sqrt{\Frequency \LowFrequency}}$, yielding the total unitary $\Unitary_\ExpandLetter$.
We can explicitly find the resulting state (see Appendix~\ref{appF}),
\begin{align}
\TwoModeGroundState_{\text{int}}  = \Unitary_\ExpandLetter \TwoModeGroundState_{\Frequency,\IntG_\Frequency}, 
\end{align}
which in the limit $\IntG_\Frequency \ll \LowFrequency^2 / \Frequency^2$ gives the fidelities
\begin{align} \Fidelity(\TwoModeGroundState_{\text{int}},\TwoModeGroundState_{\LowFrequency,\IntG_\LowFrequency}) & \approx 1 - \frac{\IntG_\Frequency^2}{32} \frac{\Frequency^4}{\LowFrequency^4}, \nonumber \\
\Fidelity(\TwoModeGroundState_{\text{int}},\TwoModeGroundState_{\LowFrequency,0}) & \approx 1 - \frac{\IntG_\Frequency^2 \pi^2}{128} \frac{\Frequency^3}{\LowFrequency^3}    , \nonumber \\
\Fidelity(\TwoModeGroundState_{\LowFrequency,\IntG_\LowFrequency},\TwoModeGroundState_{\LowFrequency,0}) & \approx 1 - \frac{\IntG_\Frequency^2}{32} \frac{\Frequency^4}{\LowFrequency^4}.
\label{fids}
\end{align}
A similar expansion protocol, based instead on an inverted harmonic potential that produces exponential delocalization, has been considered as a means of enhancing entanglement generation~\cite{Weiss2021}.
Here, by contrast, we consider parameters for which the interaction in the shallow trap becomes appreciable, $\IntG_{\LowFrequency} = \IntG_{\Frequency} \Frequency^2 / \LowFrequency^2 \lesssim 0.5$, while the expanded state $\TwoModeGroundState_{\text{int}}$ remains close to the noninteracting shallow-trap ground state $\TwoModeGroundState_{\LowFrequency,0}$.
This proximity is verified using the exact Gaussian-state fidelity rather than the weak-coupling expansion in Eqs.~\eqref{fids}.
The state $\TwoModeGroundState_{\text{int}}$ then serves as the initial state for the optimal-control step, during which the two local potentials are modulated independently according to
\begin{align}
    \frac{\Hamiltonian}{\hbar \LowFrequency} = \sum_{i=1}^2 \Big( \frac{1}{2} \MomOp_{i,\LowFrequency}^2 + \frac{1}{2} \PosOp_{i,\LowFrequency}^2 + \NonLin_{\LowFrequency} \PosOp_{i, \LowFrequency}^3 & + \force_i(\DimTime)\PosOp_{i, \LowFrequency}  \Big) \nonumber \\ 
    &+ \frac{\IntG_\LowFrequency} {2-\IntG_\LowFrequency} \PosOp_{1,\LowFrequency}\PosOp_{2,\LowFrequency},
\end{align}
We retain only the local cubic nonharmonicities of the trapping potentials.
For the symmetric transverse geometry of Equation~\eqref{cen_pot}, the central interaction is an even function of $\PositionOperator_1 - \PositionOperator_2$, so no third-order terms arise in its expansion about $\PositionOperator_1 - \PositionOperator_2 = 0$.
The leading correction beyond the quadratic interaction is fourth order and contains both local quartic terms and mixed quartic terms such as $\PositionOperator_1^3 \PositionOperator_2$, etc. These contributions are negligible for the parameters considered~\cite{Ornigotti2025}.
We use two independent controls, $\force_1(\DimTime)$ and $\force_2(\DimTime)$, to prepare one of the Bell states in the single-excitation subspace spanned by the two lowest local oscillator levels,
\begin{align}
    \ket{\pm}_\LowFrequency = \frac{1}{\sqrt{2}} \rounds{  \ket{0}_{1,\LowFrequency} \ket{1}_{2,\LowFrequency} \pm  \ket{1}_{1,\LowFrequency} \ket{0}_{2,\LowFrequency} }.
\end{align}
Here, the local Fock states are defined with respect to the harmonic parts of the individual shallow traps.
As a representative example, we consider two particles, each carrying one elementary charge and separated by $\SepD \approx 10$ {\textmu}m, using the nanoparticle and trapping parameters introduced in Sec.~\ref{sec_exp}.
These parameters correspond to a delocalization of approximately 29 dB, for which $\IntG_\Frequency \approx 8 \cdot 10^{-7}$ is enhanced to $\IntG_\LowFrequency \approx 0.5$.
We take $\NonLin_\LowFrequency = 0.04$ and $\LowFrequency \DimTimeMax = 9 \cdot 2 \pi$ for the optimal-control stage.
The inset of Fig.~\ref{Fig5}(a) shows the joint position probability density obtained in a representative optimized run.
The corresponding final state has a fidelity of approximately 96$\%$ with the target Bell state $\ket{+}$
This example demonstrates that quantum optimal control can generate non-Gaussian entanglement between two weakly nonlinear oscillators coupled through a bilinear interaction.

\subsection{Two motional modes of a single particle}
The next setup we consider involves a single particle with two active mechanical modes coupled due to the nonseparability of the realistic trapping potential [see Fig.~\ref{Fig5}(b)].
Bell states between the mechanical modes of a levitated particle provide a minimal platform for generating macroscopic, spatial-mode entanglement within a single massive object
Such states could enable probes of anisotropic decoherence and directional force sensing without requiring control of multiple particles.
Entanglement between two mechanical modes of the same physical system has been achieved in trapped ions~\cite{Jeon2024,Matsos2025,Fontbote-Schmidt2026} and quantum acoustodynamical systems~\cite{Wollack2022,vonLupke2024}.
In levitated mechanical systems, simultaneous ground-state cooling of two modes has been achieved for both translational motion~\cite{Piotrowski2023,Deplano2025} and libration~\cite{Troyer2026}.
We apply the expansion protocol simultaneously to two modes that are initially cooled to their ground states.
Then, the Hamiltonian during the control step reads
\begin{align}
    \frac{\Hamiltonian}{\hbar \LowFrequency} = \sum_{\mu \in \{\Pos,\PosY\}} \Big( \frac{1}{2} \MomOp_{\mu,\LowFrequency}^2 + \frac{1}{2} \hat{\mu}_{\LowFrequency}^2 &+ \NonLin_{\LowFrequency} \hat{\mu}_{\LowFrequency}^3  + \force_\mu(\DimTime)\hat{\mu}_{ \LowFrequency}  \Big) \nonumber \\ 
    &+ \NonLin_{\LowFrequency} ( \PosOp_{\LowFrequency}^2 \PosOpY_{\LowFrequency} + \PosOp_{\LowFrequency} \PosOpY_{\LowFrequency}^2).
\end{align}
For simplicity, we take equal strengths for the self-cubic and cross-cubic terms.
The direct mode coupling is provided by the cross-cubic terms.
When expressed in ladder operators, these terms contain three-wave-mixing processes analogous to those used in quantum optics~\cite{Boyd2008} and circuit quantum electrodynamics~\cite{Frattini2017,Sivak2019}.
We optimize two independent controls, $\force_x(\DimTime)$ and $\force_y(\DimTime)$, to transfer the shallow-trap product ground state $\TwoModeGroundState_{\Pos \PosY}$ to the antisymmetric Bell state $\ket{-}_\LowFrequency$, using the same control duration and nonharmonicity as in the two-particle example.
The inset of Fig.~\ref{Fig5}(b) shows the resulting joint position probability density.
The optimized state has a transfer fidelity of approximately 96$\%$ with the target.
This example shows that intrinsic cubic nonlinearities can be used to prepare entangled states of two mechanical modes within a single particle.

\section{Conclusions and outlook}
\label{sec_conc}
In conclusion, we have presented a protocol for preparing a broad class of target non-Gaussian states in continuous mechanical degrees of freedom of a levitated particle, including center-of-mass motion and libration.
The protocol combines experimentally accessible optimal modulation of a weakly nonharmonic potential with transient wave-function delocalization, avoiding the need for coherent coupling to auxiliary two-level systems.
A central aspect of the proposed strategy is the separation between the regimes used for nonlinear control and for storage and measurement.
Specifically, strong motional delocalization is used transiently to enhance weak intrinsic nonharmonicities, after which the final non-Gaussian state is coherently recompressed into a tightly confined trap, reducing its susceptibility to position-dependent noise and facilitating experimental readout.
Because the small zero-point spread of levitated particles suppresses their coupling to weak trapping-potential nonlinearities, we estimate the motional expansion and coherence requirements needed to prepare selected target states.
We have also shown that controlled modulation of weakly nonharmonic potentials can be extended to more complex tasks, including mechanical Bell-state preparation.
The proposed method offers a route toward universal control of the mechanical degrees of freedom of levitated massive objects.
Beyond concrete protocols targeted at levitated objects, our findings point to a broader paradigm for non-Gaussian state engineering in weakly nonlinear bosonic systems.
Quantum non-Gaussianity~\cite{Walschaers2021,Rakhubovsky2024} is a key resource for quantum metrology~\cite{Grochowski2025}, bosonic quantum information processing~\cite{Genoni2010}, and fundamental tests of quantum mechanics~\cite{Howl2021}.
However, access to this regime is often limited by weak intrinsic nonlinearities.
Here, we have demonstrated that complex non-Gaussian states can be generated in such systems without auxiliary nonlinear elements.
In fact, related strategies that enhance weak nonlinearities through state excitation~\cite{Lingenfelter2021} and combine this enhancement with optimal control~\cite{Yuan2025} have recently been proposed for cavity systems with weak Kerr nonlinearities.

In levitodynamics, the required ingredients are available across several state-of-the-art platforms, including optically trapped dielectric nanoparticles combined with electrostatic potentials~\cite{Millen2015,Conangla2020,Dania2021,Melo2024,Bonvin2024a}, magnetically levitated superconducting spheres~\cite{Hofer2023,GutierrezLatorre2023}, and librational modes with intrinsically nonharmonic potentials~\cite{Zielinska2023,Dania2025,Troyer2026}.
These platforms provide a plausible route toward versatile non-Gaussian control of levitated motional degrees of freedom.

Moreover, we emphasize that the underlying strategy is not restricted to levitated particles.
Any system in which weak intrinsic nonharmonicities can be accessed and dynamically modulated may benefit from the same control principle.
Potential applications include neutral atoms in optical potentials~\cite{Murmann2015a,Kaufman2015a,Dupont2021,Kendell2024}, trapped electrons~\cite{Osada2022}, trapped ions~\cite{Matsos2024,Matsos2025}, and Bose–Einstein condensates~\cite{Bucker2011,Bucker2013,vanFrank2016a,Ji2026}.
Beyond trapped massive systems, related control strategies may also be applied in circuit quantum electrodynamics~\cite{Eriksson2024} and quantum acoustodynamics, where weak cubic nonlinearities have recently attracted attention.
More generally, delocalization-enhanced nonlinear control may provide a broadly applicable tool for non-Gaussian bosonic state preparation and manipulation across diverse platforms.

Future directions include optimization of coherence injection in weakly nonlinear thermal systems~\cite{Yang2025}, direct optimization of certifiable non-Gaussian features and tailored witnesses~\cite{Moore2019,Zaw2023}, and hybrid nanoparticle-control protocols combining open- and closed-loop methods~\cite{Poddubny2024,Winkler2025}.
Optimal-control strategies for suppressing recoil heating constitute another promising avenue~\cite{Zhang2024}.
Further extensions involve tailoring the presented approach to interaction mechanisms beyond direct coupling, including cavity-mediated interactions~\cite{Vijayan2024}, strongly asymmetric systems with large mass ratios~\cite{Bykov2025}, and entangling protocols for few-body atomic systems that exploit larger motional Hilbert spaces for quantum information processing.
Finally, extending nonharmonicity-enhancement protocols to other mechanical platforms, including quantum-acoustic systems~\cite{Marti2024,Rahman2025} and suspended nanostructures~\cite{Chan2011}, would broaden the scope of the approach.

\begin{acknowledgments}
We acknowledge the contribution of S. Casulleras at the initial stages of the work.
We thank N. Meyer and G. Tomassi for helpful discussions.
This research has been supported by the European Research Council (ERC) under the Grant Agreement No. [951234] (Q-Xtreme ERC-2020-SyG).
P.T.G. has been supported by the project CZ.02.01.01/00/22\_010/0013054 (C-MONS) within Programme JAC MSCA Fellowships at Palacký University Olomouc IV.
The computational results presented here have been achieved in part using the LEO HPC infrastructure of the University of Innsbruck, where both authors were employed during the initial stages of the work.
We acknowledge an open-source package used for optimization, Quantum Optimal Control Suite~\cite{Rossignolo2023}.
Some of the plots have used Scientific colour maps~\cite{Crameri2020,Crameri2023}.
\end{acknowledgments}

\section*{Data availability}
Data analysis and simulation codes will be available on Zenodo~\cite{zenodo3} before the final publication.

\renewcommand{\theequation}{A\arabic{equation}}
\setcounter{equation}{0}

\appendix
\section{Averaging over optimization runs}
\label{appA}
To characterize the variability of the optimization, we perform multiple independent runs with different initial control pulses and the same computational budget.
We find that the final infidelities $1 - \Fidelity_i$ are approximately log-normally distributed.
See Fig.~\ref{FigA1} for a representative example.
We therefore use their geometric mean as a measure of typical optimization performance,
\begin{align}
    1-\AvFidelity  = \exp\squares{\frac{1}{\NoRuns} \sum_{i=1}^{\NoRuns} \log (1- \Fidelity_i)},
\end{align}
where $i$ iterates over $\NoRuns$ optimization runs.

\renewcommand{\thefigure}{A\arabic{figure}}
\setcounter{figure}{0}
\begin{figure}[b!]
    \includegraphics[width=\linewidth]{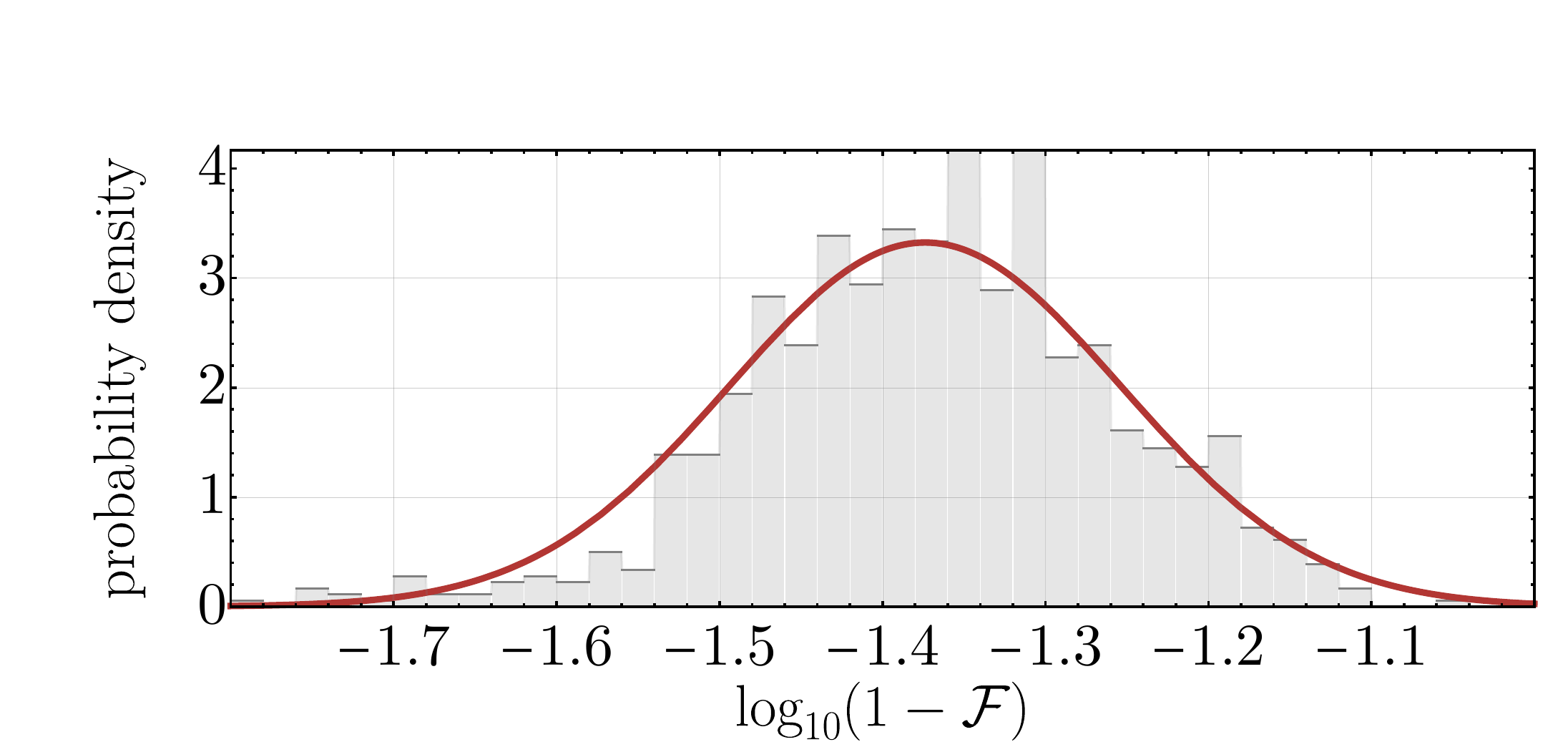}
    \caption{ Histogram of the final infidelities obtained from independent optimizations of the first excited Fock-state preparation protocol, for $\NonLin_\LowFrequency = 0.03$ and $\LowFrequency \DimTimeMax = 7 \cdot 2 \pi$. The histogram contains $\NoRuns \approx 1000$ optimization runs, each performed with the same computational budget and control constraints but different random initialization. The red curve shows a fitted log-normal distribution. Similar approximately log-normal behavior is observed for the other parameter sets considered.\label{FigA1}}
\end{figure}

\renewcommand{\theequation}{B\arabic{equation}}
\setcounter{equation}{0}

\section{Decoherence during linear dynamics}
\label{appB}
During the expansion and compression steps, the master equation~\eqref{mastereq} with $\HamiltonianCorr = 0$ can be equivalently recast onto a dynamical equation for a characteristic function $\Characteristic (\KVec,\Time)$,
\begin{equation}
    \partial_{\Time} \Characteristic (\KVec,\Time) = \rounds{\KVec \MatrixEvo \nabla_{\KVec}-2\Noise \KP^2} \Characteristic (\KVec,\Time),
\end{equation}
where $\MatrixEvo = \left[ (1- \LowFrequency / \Frequency) \PauliX - i (1+\LowFrequency / \Frequency) \PauliY \right]/2$ is expressed via Pauli matrices, $\KVec = (\KX , \KP)$, $\Time = \Frequency \DimTime$, $\Noise = \DimNoise_\ExpandLetter / \Frequency$, and
\begin{equation}
     \Characteristic (\KVec)  = \text{Tr} \left\{ \DensityMatrix \ \text{exp} \left[-i \sqrt{2} \left( \frac{\KP \MomentumOperator}{\sqrt{\hbar \Mass \Frequency}} + \frac{\KX \PositionOperator \sqrt{\Mass \Frequency}}{\sqrt{\hbar}}\right) \right]\right\}.
\end{equation}
Introducing an auxiliary characteristic function $\LiouvilleCharacteristic (\KVec,\Time) = \text{exp}(-\Time\KVec \MatrixEvo \nabla_{\KVec}) \Characteristic (\KVec,\Time)$, we can write down an evolution equation,
\begin{equation}
    \partial_{\Time} \LiouvilleCharacteristic (\KVec,\Time) = -2\Noise \KVec^{\text{T}} \MatrixNoiseAux(\Time) \KVec\LiouvilleCharacteristic (\KVec,\Time),
\end{equation}
where $\MatrixNoiseAux(\Time) = \text{exp} (-\MatrixEvo \Time) (\mathds{1} - \PauliZ) \text{exp} (-\MatrixEvo^{\text{T}} \Time)$.
It has a solution,
\begin{equation}
     \LiouvilleCharacteristic (\KVec,\Time) = e^{-\frac{1}{2} \KVec^{\text{T}} \NoiseMatrix (\tau) \KVec} \LiouvilleCharacteristic (\KVec,\Time=0),
\end{equation}
where $\NoiseMatrix (\tau) = 4 \Noise \int_0^\Time \MatrixNoiseAux(\Time') \dd \Time'$.
We want to evaluate the loss of fidelity due to decoherence, $\Fidelity = \Fidelity(\DensityMatrix_\DimNoise,\DensityMatrix_0)$, where index $\DimNoise$ corresponds to a noisy dynamics and $0$ to a noiseless one.
It can be evaluated through 
\begin{equation}
     \Fidelity = \int \dd \KVec \Characteristic_0(\KVec,\Time) \Characteristic_\DimNoise(\KVec,\Time) = \int \dd \KVec \LiouvilleCharacteristic_0(\KVec,\Time) \LiouvilleCharacteristic_\DimNoise(\KVec,\Time),
\end{equation}
where the first equality holds if one of the states is pure and follows from an analogous expression for Wigner functions~\cite{Liang2019}.
The second one follows from the identities $\LiouvilleCharacteristic(\KVec,\Time) = \Characteristic(\text{exp}(\MatrixEvo \Time) \KVec, \Time)$ and $\text{det} \ \text{exp}(\MatrixEvo \Time) = 1$.
In the limit of small $\Noise$, the fidelity can be then written as
\begin{equation}
     \Fidelity \approx 1 - \int \dd \KVec \frac{1}{2} \KVec^{\text{T}} \NoiseMatrix (\tau) \KVec \left| \LiouvilleCharacteristic (\KVec,\Time=0) \right|^2.
\end{equation}
For the expansion phase, the initial state $\LiouvilleCharacteristic (\KVec,\Time=0) $ is taken to be the ground state, while for the compression phase, it is taken to be the target state, rotated and squeezed due to previous steps of the protocol, $\LiouvilleCharacteristic (\KVec,0) = \Characteristic_\text{T} (\text{exp}(\MatrixEvo \TimeExpand) \KVec)$.
The total fidelity can then be evaluated to 
\begin{equation}
     \Fidelity \approx 1 - \DecoCoeff^\ExpandLetter \frac{\DimNoise_\ExpandLetter}{\Frequency} \left( \frac{\Frequency}{\LowFrequency}\right)^{\frac{3}{2}},
\end{equation}
where coefficient $\DecoCoeff^\ExpandLetter = \DecoCoeff^\text{ex} + \DecoCoeff^\text{cm} $ consists of the contribution from the expansion phase, $\DecoCoeff^\text{ex} = \pi / 4$, and state-dependent contribution from the compression phase, $\DecoCoeff^\text{cm}$.
For the Fock state $\ket{n}$, it reads $\DecoCoeff^\text{cm} = (2n+1) \pi / 4$, while for the cat state $\KetCatState_{\LowFrequency} \sim \ket{\CohDisLoop}_{\LowFrequency} + \ket{-\CohDisLoop}_{\LowFrequency}$ with $\CohDisLoop = \text{exp}(i \varphi) |\CohDisLoop |$ (a more general form than the one considered in the main text), it reads $\DecoCoeff^\text{cm} = (1+2 |\CohDisLoop|^2 \cos {2 \varphi} + 2 |\CohDisLoop|^2 \tanh{|\CohDisLoop|^2} ) \pi / 4$.
Here, $\varphi = 0$ corresponds to the momentum superposition at the end of the compression step, while $\varphi = \pi/2$ corresponds to the spatial one.
The latter case is the least sensitive to noise among the cat-state phase-space orientations considered.

\renewcommand{\theequation}{C\arabic{equation}}
\setcounter{equation}{0}

\section{Parameter uncertainty in expansion and compression phases}
\label{appC}
Here we consider $\DimNoise_{\ExpandLetter} = 0$ and corrections coming only from $\HamiltonianCorr$.
To characterize the effect of these corrections, let us go into the interaction picture with respect to $\Hamiltonian_{\ExpandLetter}$, such that $\HamiltonianInt (\DimTime) \equiv \Unitary_{\ExpandLetter}^{\dagger} \HamiltonianCorr (\PositionOperator) \Unitary_{\ExpandLetter} = \HamiltonianCorr [\PositionOperatorInt(\DimTime)]$, where $\Unitary_{\ExpandLetter} \equiv \text{exp}(- i \Hamiltonian_{\ExpandLetter} \DimTime / \hbar)$ and $\PositionOperatorInt (\DimTime) = \PositionOperator  \text{cos}(\sqrt{\Frequency \LowFrequency } \DimTime) + \MomentumOperator \text{sin}(\sqrt{\Frequency \LowFrequency }\DimTime) / \Mass \sqrt{\Frequency \LowFrequency } $.
We aim to compute the loss of fidelity due to $\HamiltonianCorr$, namely
\begin{equation}
    \Fidelity_j = | \expval{ \Unitary^{\text{i}}}{ \WaveFunction_j^{\text{i}}} |^2, 
\end{equation}
where $j \in \{\text{ex},\text{cm}\}$ corresponds to expansion and compression steps, $\ket{\WaveFunction_\text{ex}} = \ket{0}_{\Frequency}$, $\ket{\WaveFunction_\text{cm}} = \ket{\TargetState}_{\LowFrequency}$, and the unitary evolution $\Unitary^{\text{i}}$ can be perturbatively approximated via second-order Dyson series,
\begin{align}
     \Unitary^{\text{i}} \approx 1 &- \frac{i}{\hbar} \int_0^{\DimTimeExpand} \dd \DimTime' \HamiltonianInt (\DimTime') \nonumber \\ &+ \left(\frac{i}{\hbar}\right)^2 \int_0^{\DimTimeExpand} \dd \DimTime' \int_0^{\DimTime'}  \dd \DimTime'' \HamiltonianInt (\DimTime') \HamiltonianInt (\DimTime'').
\end{align}
After explicit calculation, one gets the total loss of fidelity, $\Fidelity \approx \Fidelity_\text{ex} \Fidelity_\text{cm}$, and, consequently, $1 - \Fidelity \approx (1 - \Fidelity_\text{ex}) + (1 -\Fidelity_\text{cm})$.
In the leading order,
\begin{align}
     1 - &\Fidelity = \nonumber \\ & \ForceCoeff^{\ExpandLetter} \left( \frac{\corrForce}{\Force_\Frequency} \frac{\Frequency}{\LowFrequency}\right)^2 +  \HarmonicCoeff^{\ExpandLetter}  \left[ \frac{\corrFreq}{\Frequency} \left(\frac{\Frequency}{\LowFrequency}\right)^{\frac{3}{4}}\right]^4 +  \CubicCoeff^{\ExpandLetter} \left[ \frac{\corrNonlin}{\CubNonlin_\Frequency} \left(\frac{\Frequency}{\LowFrequency}\right)^{2}\right]^2,
\end{align}
where $\Force_\Frequency = \hbar \Frequency / \PosDim_{\Frequency}$, $\CubNonlin_\Frequency = \hbar \Frequency / \PosDim_{\Frequency}^3$, and $\DecoCoeffMain^{\ExpandLetter}_i$ have contributions from both expansion and compression steps, $\DecoCoeffMain_j^{\ExpandLetter} = \DecoCoeffMain_j^{\text{ex}} + \DecoCoeffMain_j^{\text{cm}}$.
The former can be explicitly evaluated as $\DecoCoeffMain_2^{\text{ex}} = 1/2$, $\DecoCoeffMain_3^{\text{ex}} = 9 \pi^2 / 768$, $\DecoCoeffMain_4^{\text{ex}} = 5/6$, while the latter are target-state-dependent and can be evaluated numerically. 
In Fig.~\ref{FigC1}, these coefficients are plotted for Fock and cat states.
Again, the latter case is the least sensitive to the fidelity loss due to parameter uncertainty between all the cat state phase-space orientations.

\renewcommand{\thefigure}{C\arabic{figure}}
\setcounter{figure}{0}

\begin{figure}[t!]
    \includegraphics[width=\linewidth]{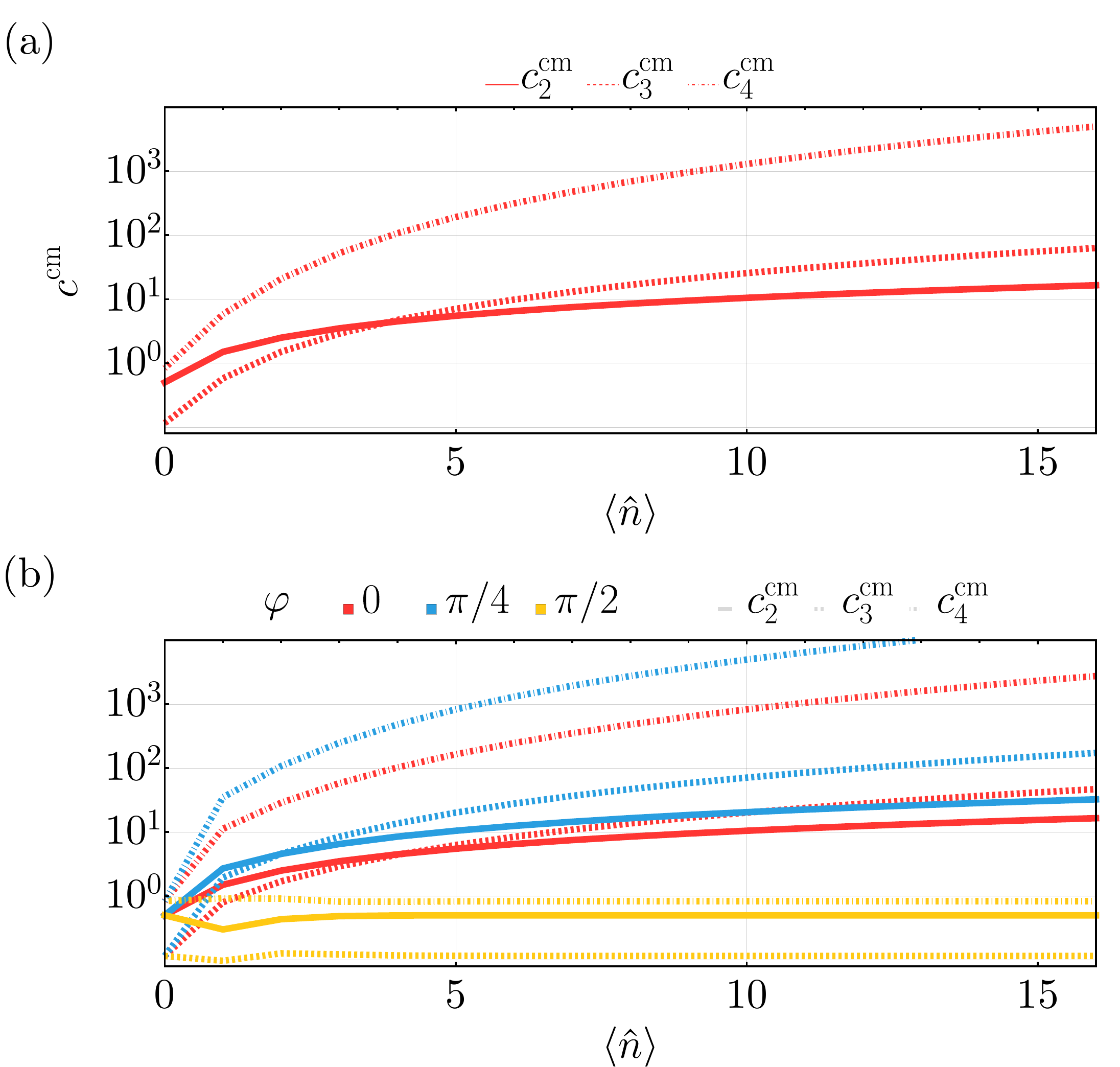}
    \caption{ Coefficients $\DecoCoeffMain_j^{\text{cm}}$ for (a) Fock states and (b) cat states.\label{FigC1}}
\end{figure}

\renewcommand{\thefigure}{E\arabic{figure}}
\setcounter{figure}{0}
\begin{figure*}[ht!]
    \includegraphics[width= \linewidth]{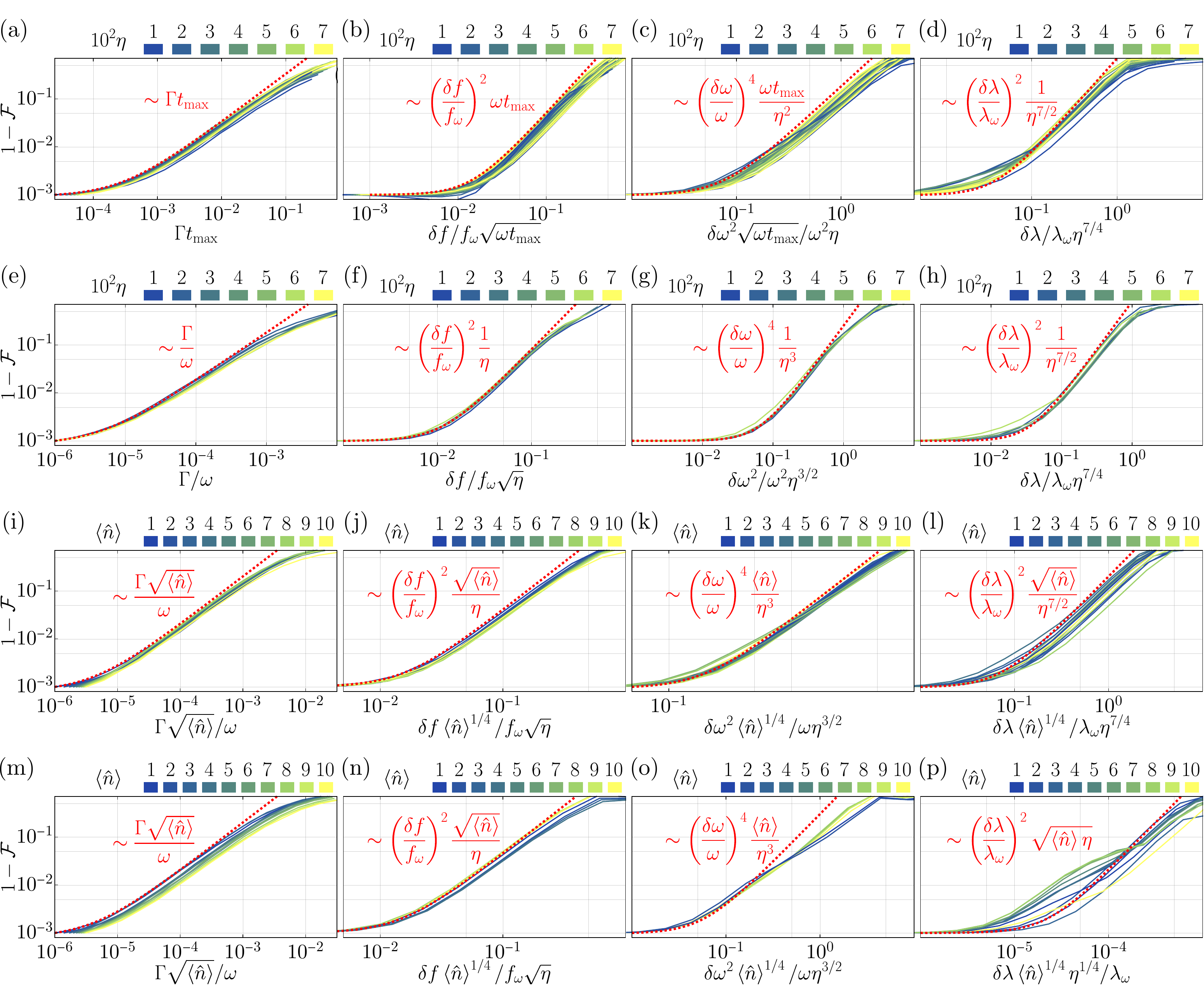}
    \caption{Data corresponding to fits in Tab.~\ref{tabnum}. Columns from left to right correspond to different sources of noise defined in~\eqref{mastereq}: mechanical diffusion and linear, quadratic, and cubic deviations from the ideal Hamiltonian. Rows from top to bottom correspond to the first excited Fock state preparation (all cases), the first excited Fock state preparation (minimized over the duration of the protocol), large Fock state preparation, and cat state preparation.
    \label{FigE1}}
\end{figure*}

\renewcommand{\theequation}{D\arabic{equation}}
\setcounter{equation}{0}

\section{Numerical details}
\label{appD}
For all the simulations, the split-step method~\cite{Leforestier1991a} was used to simulate the dynamics on a spatial grid.
The numerical parameters read: number of spatial grid points $N_x = 2^{10}$, where the grid spanned $x \in [-26,26]$, and number of time grid points $N_\Time = 1000$.
After optimizing the pulses, the dynamics was run with increased accuracy to make sure the results converged.
The optimization has been performed with QuoCS library version 0.0.44~\cite{Rossignolo2023}.
The number of Fourier components was 50, and they were randomized from a uniform distribution on a fixed bandwidth.
It was checked that increasing the bandwidth does not change the results.
Each optimization procedure involved 80 independent optimization runs with a fixed physical run time (60 hours of single-core wall-clock time) on a single-core architecture.
We solved noisy dynamics by averaging over different noise realizations and constructing a Monte Carlo density matrix~\cite{Molmer1993}.
For each of noise realizations, we employed stochastic fluctuations of the linear term, $\DimForce(\DimTime) \rightarrow \DimForce(\DimTime) + \DimForce_\Fluct(\DimTime)$, where $\DimForce_\Fluct(\DimTime)$ is a stochastic force of zero mean and assumed delta-correlated in the relevant time scales, namely, $\left\langle \DimForce_\Fluct(\DimTime) \DimForce_\Fluct(\DimTime') \right\rangle = 2 \hbar^2 \DimNoise_{\ShakeLetter} \delta(\DimTime-\DimTime') / \PosDim_{\Frequency}^2$.

\begin{table}[t!]
\begin{tabular}{c||c||c|c|c|c|c|c}
\hline  \hline
          Case        & $j$ & $c_j^{\ShakeLetter}$ & $\alpha_j$ &  $\beta_j$ & $\gamma_j$ & $\delta_j$ & $\epsilon_j$   \\ \hline \hline
\multirow{4}{*}{(\textsc{a}) 1$^{\text{st}}$ Fock all cases } & 1 & 3.3 & 1 & - & 1 & - & 1 \\ \cline{2-8} 
                  & 2 & 5.0 & 2 & - & 2 & - & 1 \\ \cline{2-8} 
                  & 3  & 0.18 & 4 & -2 & 3 & - & 1 \\ \cline{2-8}               
                  & 4 & 0.7 & 2 & -7/2 & 5 & - & - \\ \hline \hline
\multirow{4}{*}{(\textsc{b}) 1$^{\text{st}}$ Fock min. inf.} & 1 & 180 & 1 & - & 2 & - & - \\ \cline{2-8} 
                  & 2 & 8.0 & 2 & -1 & 3 & - & - \\ \cline{2-8} 
                  & 3 & 0.25 & 4 & -3 & 4 & - & - \\ \cline{2-8}                  
                  & 4 & 0.8 & 2 & -7/2 & 5 & - & - \\ \hline \hline
\multirow{4}{*}{(\textsc{c}) High Fock} & 1 & 200 & 1 & - & 2 & 1/2 & - \\ \cline{2-8} 
                  & 2 & 4.0 & 2 & -1 & 3 & 1/2 & - \\ \cline{2-8} 
                  & 3 & 0.025 & 4 & -3 & 4 & 1/2 & - \\ \cline{2-8}                   
                  & 4 & 0.2 & 2 & -7/2 & 5 & 1/2 & - \\ \hline \hline
\multirow{4}{*}{(\textsc{d}) Cat} & 1 & 200 & 1 & - & 2 & 1/2 & - \\ \cline{2-8} 
                  & 2 & 4.5 & 2 & -1 & 3 & 1/2 & - \\ \cline{2-8} 
                  & 3 & 0.3 & 4 & -3 & 4 & 1/2 & - \\ \cline{2-8}                    
                  & 4 & 2$\cdot10^6$ & 2 & 1/2 & 5 & 1/2 & - \\ \hline \hline
\end{tabular}
\caption{Fidelity loss coefficients given by~\eqref{numfit} during the optimal control step for different sources of noise. For details see Appendix~\ref{appE}.}
\label{tabnum}
\end{table}

\renewcommand{\theequation}{E\arabic{equation}}
\setcounter{equation}{0}

\section{Summary of fidelity loss during optimal control step}
\label{appE}
The fidelity loss during nonlinear dynamics of the optimal control step needs to be evaluated numerically. 
Here we report the fits for the upper bounds for these fidelity losses for four considered cases: (\textsc{a}) preparation of the first Fock state, (\textsc{b}) the same but protocols are chosen to minimize the infidelity at a given $\DimNoise_\ShakeLetter$ (cf. the main text), (\textsc{c}) preparation of high Fock states, (\textsc{d}) preparation of cat states.
In each case, the fidelity loss is expressed as
\begin{align}
    1-\Fidelity_j = c_j^{\ShakeLetter} \delta g_j^{\alpha_j} \NonLin_{\LowFrequency}^{\beta_j} \left(\frac{\Frequency}{\LowFrequency}\right)^{\gamma_j} \left<\hat{n}\right>^{\delta_j} \left( \Frequency \DimTimeMax \right)^{\epsilon_j},
    \label{numfit}
\end{align}
where $j \in \{1,\dots,4\}$, $\delta g \in \{ \DimNoise_\ShakeLetter / \Frequency, \corrForce/ \Force_\Frequency, \corrFreq / \Frequency, \corrNonlin /  \CubNonlin_\Frequency\}$ correspond to different sources of noise defined in~\eqref{mastereq}.
Values of constants in~\eqref{numfit} are fitted such that all the fidelity loss results for a given case fall into one universal curve after offsetting by an initial, coherent value of the infidelity [cf. Figs.~\ref{Fig4}(b)-(d)].
These values are presented in Tab.~\ref{tabnum}.
The visual representation of the results and fits is placed in Fig.~\ref{FigE1}.

\renewcommand{\theequation}{F\arabic{equation}}
\setcounter{equation}{0}

\section{The expansion protocol with two interacting particles}
\label{appF}
The bilinearized interacting Hamiltonian~\eqref{linearized} can be diagonalized by introducing relative and center-of-mass motion of two particles, $\PositionOperator_\mp = (\PositionOperator_1 \mp \PositionOperator_2) /\sqrt{2}$,
\begin{align}
    \Hamiltonian = \sum_{j=\pm} \frac{\PositionOperator^2_\pm}{2 \Mass} + \frac{1}{2} \Mass \UnFreq_\pm \PositionOperator^2_\pm,
    \label{ham_rel_com}
\end{align}
where $\UnFreq_+ = \UnFreq$ and $\UnFreq_- = \UnFreq \sqrt{1-\IntG_\UnFreq}$.
The starting state of the expansion protocol is the ground state of the harmonic oscillator~\eqref{ham_rel_com} with $\UnFreq_+ = \Frequency$ and $\UnFreq_- = \Frequency \sqrt{1-\IntG_\Frequency}$, which is then quenched to frequencies $\UnFreq_+ = \sqrt{\Frequency \LowFrequency}$ and $\UnFreq_- = \sqrt{\Frequency \LowFrequency}\sqrt{1-\IntG_{\sqrt{\Frequency \LowFrequency}}}$ and evolved for time $\DimTimeExpand = \pi / 2 \sqrt{\Frequency \LowFrequency}$.
For the center-of-mass motion, such a protocol yields a ground state of the harmonic oscillator with frequency $\LowFrequency$, while for the relative one, the state remains zero-mean Gaussian that can be written in position basis as
\begin{align}
    \WaveFunction(\PosDim,\DimTime) = \left[ \frac{\Mass \text{Re} A(\DimTime)}{\pi \hbar}\right]^{1/4} e^{-\frac{m A(\DimTime) \PosDim^2 }{2 \hbar}}.
    \label{Ansatz}
\end{align}
The evolution in a harmonic potential with frequency $\UnFreq$ can be solved via substituting Ansatz~\eqref{Ansatz} into the Schrödinger equation, yielding
\begin{align}
    \ImagUnit \PhaseFunction'(\DimTime) = \PhaseFunction^2(\DimTime) - \UnFreq^2, \ \ 2 \text{Im}[\PhaseFunction(\DimTime)] \text{Re}[\PhaseFunction(\DimTime)] = \text{Re}[\PhaseFunction'(\DimTime)],
\end{align}
which can be solved as
\begin{align}
    \PhaseFunction (\DimTime) = \ImagUnit \UnFreq \tan \left[ \UnFreq \DimTime - \ImagUnit \ \text{arctanh}\rounds{\frac{\PhaseFunction_0}{\UnFreq}} \right]
\end{align}
with $\PhaseFunction(\DimTime = 0) = \PhaseFunction_0$.
Then, the state of the relative motion after the expansion protocol can be explicitly written as
which can be solved as
\begin{align}
    \PhaseFunction_{\text{int}} = \ImagUnit \Frequency \sqrt{\FrRatio - \IntG_\Frequency} \tan \left[ \frac{\pi}{2} \sqrt{1-\frac{\IntG}{\FrRatio}} - \ImagUnit \ \text{arctanh}\rounds{\sqrt{\frac{\FrRatio - \IntG_\Frequency}{1 - \IntG_\Frequency}}} \right]
\end{align}
with $\FrRatio = \LowFrequency / \Frequency$.
The initial ground state is given by $\PhaseFunction_{\Frequency} = \Frequency \sqrt{1 - \IntG_\Frequency}$, the ideal interacting ground state in the shallow nonoptical potential by $\PhaseFunction_{\LowFrequency} = \Frequency \FrRatio \sqrt{1 - \IntG_\Frequency/\FrRatio^2}$.
The fidelity between Gaussian states can be readily derived from~\eqref{Ansatz},
\begin{align}
    \Fidelity(\PhaseFunction_1,\PhaseFunction_2) = \frac{2 \sqrt{\text{Re} \PhaseFunction_1 \ \text{Re} \PhaseFunction_2  }}{| \PhaseFunction_1^* + \PhaseFunction_2 |},
\end{align}
which allows us to directly compute Eqs.~\eqref{fids}.

\nocite{Merkel2009}
\bibliography{FockLoop}
\end{document}